\documentclass[sigconf, screen]{acmart}

\AtBeginDocument{%
  }

\setcopyright{none} 
\copyrightyear{2026}
\acmYear{2026}
\acmDOI{XXXXXXX.XXXXXXX}

\acmISBN{978-X-XXXX-XXXX-X/XX/XX}

\usepackage{subcaption}
\usepackage{multirow}
\usepackage{booktabs}
\usepackage[ruled]{algorithm2e}
\usepackage{booktabs}   
\usepackage{tabularx}   
\usepackage{array}  

\begin{document}

\title{ARGON: A GNN-Empowered Compilation Framework for Scalable Neutral Atom Computing}


\author{Wenjie Sun}
\email{202211022528@std.uestc.edu.cn}
\affiliation{%
	\institution{University of Electronic Science and Technology of China}
	\country{China}
}

\author{Xiaoyu Li}
\authornote{Corresponding author.}
\email{xiaoyuuestc@uestc.edu.cn}
\affiliation{%
	\institution{University of Electronic Science and Technology of China}
	\country{China}
}

\author{Zhigang Wang}
\email{zhigangwang@uestc.edu.cn}
\affiliation{%
	\institution{University of Electronic Science and Technology of China}
	\country{China}
}

\author{Lianhui Yu}
\email{202222120303@std.uestc.edu.cn}
\affiliation{%
	\institution{University of Electronic Science and Technology of China}
	\country{China}
}

\author{Geng Chen}
\email{202312081614@std.uestc.edu.cn}
\affiliation{%
	\institution{University of Electronic Science and Technology of China}
	\country{China}
}

\author{Guowu Yang}
\email{guowu@uestc.edu.cn}
\affiliation{%
	\institution{School of Computer Science and Engineering, University of Electronic Science and Technology of China}
	\country{China}
}

\renewcommand{\shortauthors}{Sun, et al.}



%%%%%% -- PAPER CONTENT STARTS-- %%%%%%%%

\begin{abstract}
	Neutral atom quantum systems provide a promising pathway toward large-scale quantum computing owing to their high qubit uniformity and flexible connectivity. To fully exploit this architecture, compilers must strategically coordinate considerations of dynamic atom transport alongside the execution of highly parallel entangling gates. As the scale of quantum circuits increases, the intricate interplay between these operations becomes an unavoidable bottleneck in the system stack, as larger circuits introduce denser logical interactions and longer temporal dependencies. This requires the compiler to simultaneously satisfy rigid spatial interference constraints and increasingly complex movement schedules. Existing compilation methodologies typically resolve these dimensions jointly, resulting in an exponentially expanding search space that incurs substantial compilation overheads or compromises physical execution fidelity as circuit size grows.
	
	In this work, we propose ARGON, a scalable compilation framework that introduces a spatiotemporal decoupling paradigm for neutral atom processors. Our key novelty lies in offloading the static geometric conflict resolution to an offline phase, precomputing a discrete library of hardware-certified, high-parallelism spatial layouts. To guide the temporal routing process, we deploy an Graph Neural Network (GNN) predictor. This enables the compiler to evaluate candidate layouts against deep temporal horizons, proactively anticipating and evading costly downstream kinematic bottlenecks to ensure global routing efficiency. Finally, a heuristic router is developed to translate the selected sequence into collision-free physical transport.
	
	Evaluations across varied benchmark scales show that ARGON completes compilation in under 10 seconds, delivering up to a $>10^4\times$ speedup and an average speedup above $600\times$ compared to state-of-the-art baselines. With these improvements, ARGON still minimizes cumulative routing decoherence and significantly reduces overall Rydberg stages, improving execution fidelity by up to a $10^2\times$ factor on dense quantum circuits.
\end{abstract}
\begin{CCSXML}
	<ccs2012>
	<concept>
	<concept_id>10010583.10010786.10010813.10011726</concept_id>
	<concept_desc>Hardware~Quantum computation</concept_desc>
	<concept_significance>500</concept_significance>
	</concept>
	<concept>
	<concept_id>10010520.10010521.10010542.10010550</concept_id>
	<concept_desc>Computer systems organization~Quantum computing</concept_desc>
	<concept_significance>500</concept_significance>
	</concept>
	<concept>
	<concept_id>10010147.10010178</concept_id>
	<concept_desc>Computing methodologies~Artificial intelligence</concept_desc>
	<concept_significance>500</concept_significance>
	</concept>
	</ccs2012>
\end{CCSXML}

\ccsdesc[500]{Hardware~Quantum computation}
\ccsdesc[500]{Computer systems organization~Quantum computing}
\ccsdesc[500]{Computing methodologies~Artificial intelligence}

\keywords{Quantum Computing, Quantum Compilation, Graph Neural Networks, Neutral Atom Computing}

\maketitle

\section{Introduction}

Quantum computing is expected to enable breakthroughs in solving problems that are computationally prohibitive for classical systems, spanning domains such as cryptography, optimization, and quantum simulation~\cite{Arute2019,Bernstein2017,Harrigan2021,Lancellotti2024,Abbas2024,Clinton2024}. As physical implementations progress toward larger and more reliable architectures, neutral atom (NA) architectures offer compelling microarchitectural solutions to the fundamental scaling and operational bottlenecks of quantum processing~\cite{Ebadi2022,Bluvstein2022,Evered2023,Bluvstein2024}. Unlike solid-state modalities, individual neutral atoms provide strictly identical qubit properties, eliminating localized calibration anomalies and ensuring processor-wide uniformity~\cite{Henriet2020,Singh2022}. This intrinsic homogeneity enables the realization of massive qubit arrays with exceptionally long coherence times~\cite{Covey2023}. Furthermore, a defining architectural shift in modern NA platforms is the introduction of dynamically reconfigurable atom arrays \cite{Bluvstein2022}. By utilizing optical tweezers to physically reposition qubits on neutral atom arrays, this architecture natively supports extended interaction connectivities and the highly parallel execution of non-local multi-qubit logic gates.

This architectural evolution from fixed-topology grids to dynamically reconfigurable arrays fundamentally shifts the burden of quantum compilation from pure logic synthesis to complex spatiotemporal microarchitectural scheduling. In early hardware constrained by fixed quantum register pairs, compilation primarily focused on inserting deep, decoherence-prone SWAP sequences to realize long-range interactions \cite{Baker2021, Patel2022, Graham2022, Ebadi2022, Patel2023, Li2023}. However, the advent of optical tweezer-driven mobility \cite{Bluvstein2022} introduces a new paradigm: the compiler must now directly orchestrate physical atom transport. Recent advancements have successfully leveraged this mobility to achieve high-fidelity parallel entangling gates \cite{Evered2023}, logical qubit operations \cite{Bluvstein2024}, efficient error correction \cite{Xu2024}, fault-tolerant interconnects \cite{Sinclair2025}, and scaling to thousands of qubits \cite{Chiu2025, Bluvstein2025}. 

Motivated by these hardware capabilities, compilation techniques tailored for reconfigurable atom arrays have steadily advanced \cite{Tan2022, Tan2024, Wang2024, Ludmir2024, Huang2025, Tan2025}. Despite this progress, achieving computationally efficient and scalable compilation while simultaneously optimizing dynamic Acousto-Optic Deflector (AOD)-mediated qubit transport and highly parallel gate execution remains a formidable challenge \cite{Tan2022, Tan2024, Huang2025}. For instance, frameworks relying on stochastic or memoryless search strategies face inherent scalability limitations when navigating dense processor layouts \cite{Ludmir2024, Tan2025}, whereas others revert to auxiliary SWAP operations to resolve routing conflicts, inadvertently reintroducing fidelity degradation \cite{Wang2024}.

These algorithmic challenges do not stem from limitations in method design, but rather from the intrinsic difficulty of the microarchitectural scheduling problem itself. Orchestrating atom transport introduces a highly complex bottleneck, as it inextricably links the physical positioning of qubits with the time-evolving dataflow of the circuit. Specifically, to orchestrate continuous execution, the compiler is forced to navigate an intricate trade-off between two interdependent physical realities. First, it must satisfy \textbf{static spatial constraints}, which dictate that parallel Rydberg gates maintain strict interaction radius and disjoint exclusion zones to prevent crosstalk. Second, it must resolve \textbf{temporal dependencies}, which involve calculating kinematic, collision-free routing paths for continuous atom shuttling across sequential circuit layers. A spatially optimal placement for the current operation often necessitates excessive or extended temporal routing in subsequent steps, making these two dimensions tightly entangled.

\begin{figure}[t]
	\centering
	\begin{subfigure}{0.5\columnwidth}
		\centering
		\includegraphics[width=\linewidth]{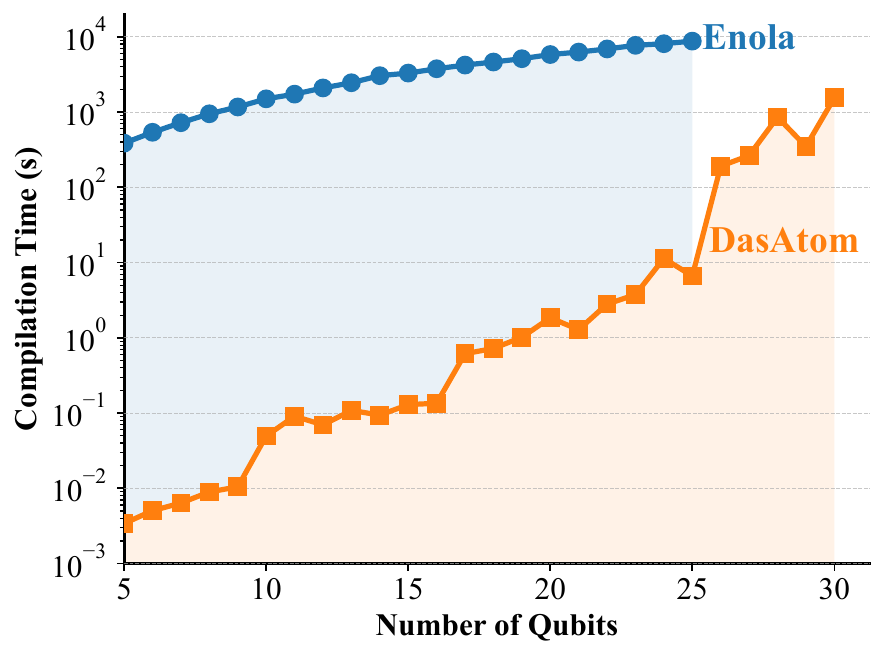} 
		\caption{Compilation time scale-up.}
		\label{fig:intro_time}
	\end{subfigure}
	\begin{subfigure}{0.49\columnwidth}
		\centering
		\includegraphics[width=\linewidth]{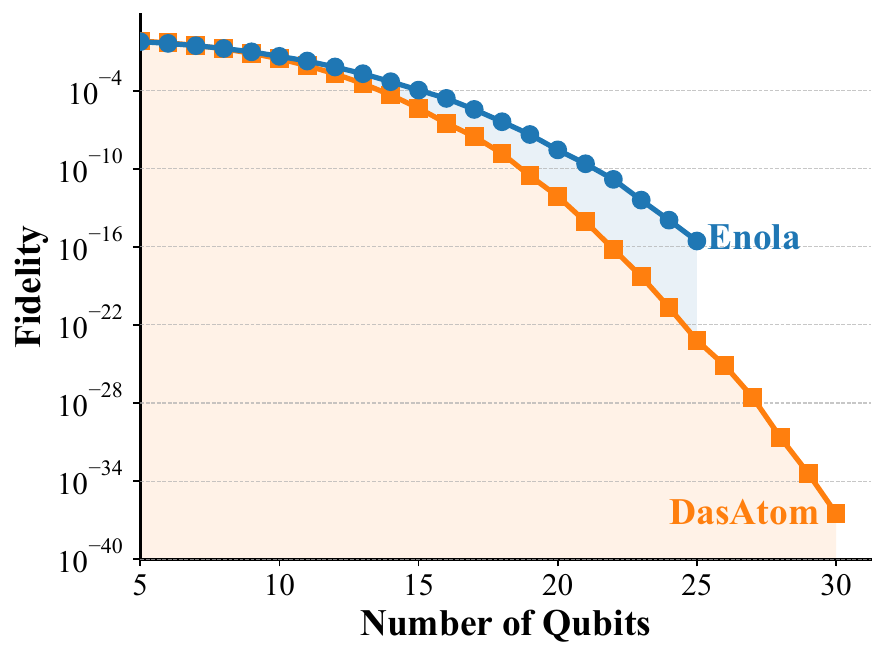}
		\caption{Fidelity degradation.}
		\label{fig:intro_fid}
	\end{subfigure}
	\caption{The scalability bottleneck of SOTA NA compilers.}
	\label{fig:intro}
\end{figure}

Existing state-of-the-art compilation methodologies typically attempt to evaluate these static and dynamic dimensions concurrently. While this joint optimization is effective for small-scale demonstrations, we observe that it inadvertently generates an exponentially expanding search space as circuit depth and array dimensions increase. Consequently, as the system scales toward practical utility, this tight spatiotemporal coupling becomes the root cause of a devastating scalability wall. 

This structural limitation is empirically validated in Fig.~\ref{fig:intro} using the random gates benchmark from MQT Bench~\cite{Quetschlich2023}. As illustrated in Fig.~\ref{fig:intro_time}, the compilation time of state-of-the-art joint-optimization solvers (e.g., Enola~\cite{Tan2025}) increases exponentially, hitting a $10^4$-second timeout limit at merely $N=25$ qubits---a multi-hour latency that inherently precludes the rapid adaptation required for practical quantum compilation. Conversely, heuristic approaches (e.g., DasAtom~\cite{Huang2025}) that prune the search space to maintain tractability experience a severe fidelity degradation. As shown in Fig.~\ref{fig:intro_fid}, the physical execution fidelity of DasAtom drops to $10^{-37}$ at $N=30$ qubits, rendering the computation impractical for physical execution. This evidence highlights a fundamental microarchitectural dilemma: forcing the compiler to jointly resolve static geometric interference and routing at compile-time dictates a critical trade-off between excessive compilation latency and substantial execution fidelity loss.

To overcome this structural scalability limitation, we propose ARGON, a highly scalable compilation framework built on a predictive spatiotemporal decoupling paradigm. Instead of jointly optimizing physical feasibility and topological evolution during compilation, ARGON isolates static spatial verification from temporal routing. This decoupling enables the compiler to elegantly handle each dimension independently, before ultimately synthesizing them back into a cohesive, executable solution. The framework realizes this philosophy through three complementary phases:

\begin{itemize}
	\item \textbf{Offline Spatial Layout Library Construction.} We recognize that physical hardware interference rules are static and entirely independent of the dynamically changing quantum circuit. By extracting spatial conflict resolution completely out of the critical compilation loop, ARGON formalizes the required Rydberg stages offline. This phase acts as a structural filter, proactively shielding the compiler from the combinatorial explosion while guaranteeing gate parallelism.
	
	\item \textbf{Graph Neural Network-Guided Predictive Placement.} With static spatial constraints resolved offline, the compiler focuses exclusively on minimizing routing overhead. Instead of relying on myopic heuristics that often lead to downstream kinematic conflicts, ARGON employs a Graph Neural Network (GNN) to capture both the hardware interaction topology and future logical dependencies. This predictive policy provides long-range topological foresight, proactively mitigating routing congestion through efficient constant-time neural inference.
	
	\item \textbf{Heuristic Parallel Routing.} In the final phase, a lightweight backend router translates the selected layouts into coordinated, collision-free AOD transport sequences. By isolating the expensive spatiotemporal decisions in the preceding stages, routing is reduced to efficient kinematic pathfinding with lightweight deadlock resolution and parallel scheduling, enabling rapid instruction synthesis with minimal overhead.
\end{itemize}

Through this decoupled architecture, ARGON successfully transforms the intractable compilation bottleneck into a highly efficient predictive inference task. Our main contributions are summarized as follows:

\begin{enumerate}
	\item We propose ARGON, a neural-predictive compilation framework for neutral atom processors that breaks the intractable scheduling bottleneck by decoupling static spatial hardware constraints from temporal routing.
	\item To operationalize this decoupling paradigm, ARGON integrates an offline SMT-constructed spatial layout library to guarantee high gate parallelism, a GNN-based predictor to evade kinematic conflict, and a lightweight heuristic router for efficient physical synthesis.
	\item Evaluations demonstrate that ARGON achieves deterministic rapid responsiveness in under 10 seconds—delivering up to a $>600\times$ speedup over state-of-the-art baselines—while simultaneously improving execution fidelity by up to a $10^2\times$ factor.
\end{enumerate}

\section{Background and Motivation}

\subsection{Hardware Primitives and Operations}

Unlike conventional solid-state platforms constrained by fixed physical wiring, neutral atom architectures confine and manipulate identical qubits in a vacuum using programmable optical tweezers~\cite{Bluvstein2022}. This architecture relies on two complementary optical systems to establish its reconfigurable geometry: a Spatial Light Modulator (SLM) that generates a dense 2D grid of stationary traps, and an Acousto-Optic Deflector (AOD) that provides a mobile array of traps~\cite{Tan2022, Huang2025}. During computation, qubits can be dynamically transferred between the static SLM and the mobile AOD traps~\cite{Tan2024}. By physically transporting atoms across the processor, this mobility fundamentally transforms the hardware into a dynamically field-programmable qubit array. It enables flexible, non-local interaction connectivities without relying on the deep, decoherence-prone SWAP gate sequences required by fixed nearest-neighbor architectures~\cite{Ebadi2022, Evered2023}.

With the microarchitectural geometry established by the optical traps, logical quantum operations are executed via precisely orchestrated laser pulses. Single-qubit operations are realized through local Raman transitions, providing high-fidelity, individually addressable state rotations. Conversely, multi-qubit entangling operations, such as controlled-Z (CZ) gates, are driven by the Rydberg blockade mechanism~\cite{Bluvstein2024, Evered2023}. By exposing the array to a global Rydberg laser pulse, neutral atoms situated within a specific proximity undergo strong dipole-dipole interactions, thereby generating quantum entanglement. Crucially, this mechanism dictates that the execution of entangling gates is strictly bound to the physical distance between participating qubits. This mechanism dictates that logical gate execution is strictly bound to the physical proximity of qubits. Consequently, the logical evaluation of the quantum circuit is inextricably tied to the spatial packing of the processor, imposing a fundamental architectural dependency that necessitates joint optimization of placement and scheduling.

\begin{figure*}[t]
	\centering
	\begin{subfigure}{0.32\textwidth}
		\centering
		\includegraphics[width=\linewidth, trim=18cm 10.5cm 18cm 10.5cm, clip]{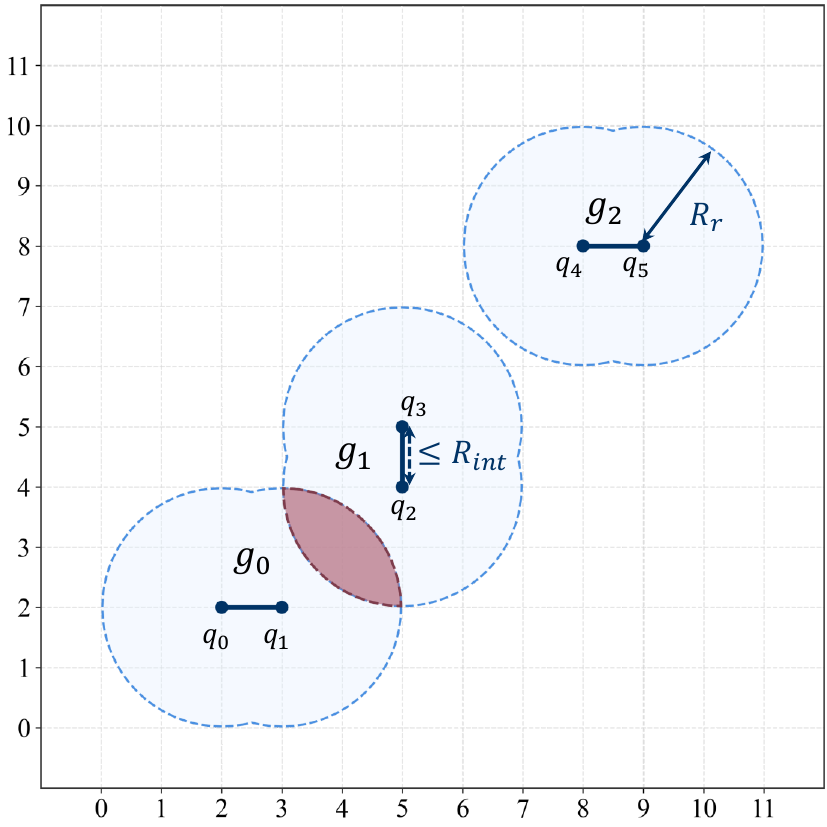}
		\caption{Proximity \& parallel execution constraint.}
		\label{fig:fig1-premi1}
	\end{subfigure}
	\hfill
	\begin{subfigure}{0.32\textwidth}
		\centering
		\includegraphics[width=\linewidth, trim=18cm 10.5cm 18cm 10.5cm, clip]{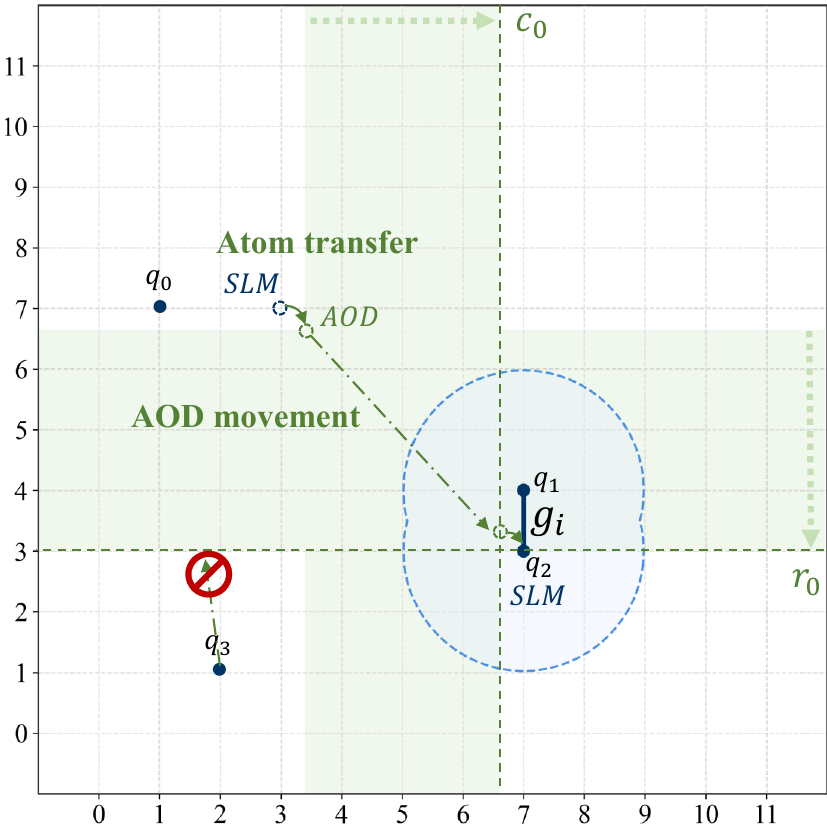}
		\caption{Cycle 1: transfer and AOD movement.}
		\label{fig:fig1-premi2}
	\end{subfigure}
	\hfill
	\begin{subfigure}{0.32\textwidth}
		\centering
		\includegraphics[width=\linewidth, trim=18cm 10.5cm 18cm 10.5cm, clip]{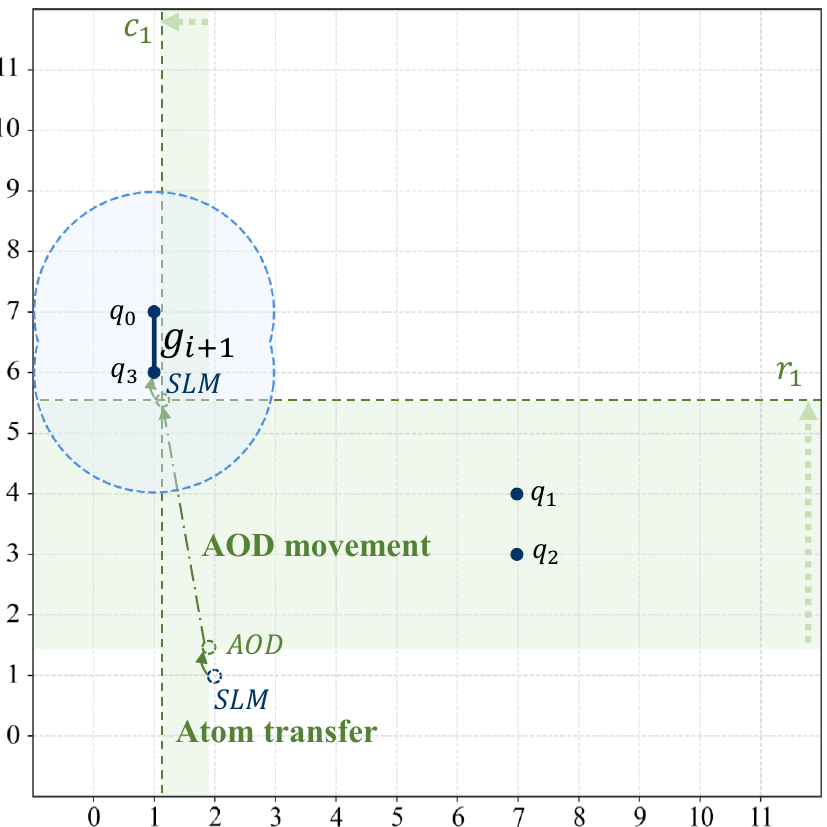}
		\caption{Cycle 2: deferred AOD movement.}
		\label{fig:fig1-premi3}
	\end{subfigure}
	\caption{Spatiotemporal scheduling constraints in reconfigurable neutral atom arrays. Blue regions represent conflict regions formed during two-qubit gate execution; for parallel execution, these regions must not overlap. Green dashed lines indicate AOD columns $c_i$ and rows $r_j$, and green areas represent regions swept by AOD movement. (a) Illustration of conflict regions for two-qubit gates and parallel execution constraints. (b)(c) Sequential execution of atom transfer and AOD movement across two steps, where two-qubit gates $g_i$ and $g_{i+1}$ cannot be executed simultaneously due to AOD movement constraints.}
	\label{fig:fig1-preliminary}
\end{figure*}

\subsection{Spatiotemporal Scheduling Constraints}

To successfully map quantum circuits onto this reconfigurable hardware, the compiler must function as a sophisticated microarchitectural scheduler, navigating two stringent and fundamentally intertwined physical dimensions.

\textbf{The first dimension involves static spatial constraints}, which dictate the strict physical boundaries of Instruction-Level Parallelism (ILP) on the processor. As established, enabling two-qubit entanglement requires participating atoms to be placed within a maximum interaction radius ($R_{int}$). Furthermore, to suppress hardware crosstalk during parallel execution, each active gate generates a spatial exclusion zone defined by a restriction radius ($R_r$, typically $R_r \ge R_{int}$)~\cite{Huang2025}. In microarchitectural terms, these disjoint geometric zones act as structural hazards: any overlap between the exclusion zones of concurrent gates implies a resource conflict, forcing the scheduler to serialize their execution.

This ILP limitation is illustrated in Fig.~\ref{fig:fig1-preliminary}(a), where three pending gates---$g_0(q_0, q_1)$, $g_1(q_2, q_3)$, and $g_2(q_4, q_5)$---compete for execution. Because the Rydberg conflict regions (red areas) of $g_0$ and $g_1$ intersect, they trigger a structural hazard and cannot be issued in the same cycle. Conversely, $g_2$ maintains an isolated spatial footprint, allowing it to be seamlessly scheduled in parallel with either $g_0$ or $g_1$, thereby fully exploiting the available ILP under the hardware's geometric limits.

Beyond maximizing static parallelism, the compiler must simultaneously navigate \textbf{the second dimension: dynamic kinematic constraints}, which govern the physical transport of qubits across the array. In reconfigurable architectures, AOD traps operate under strict mobility rules: atoms sharing an AOD row or column must move in tandem, and active rows or columns can never cross one another during transit to prevent optical interference and atom loss~\cite{Tan2022, Huang2025}. Architecturally, these physical limitations manifest as kinematic routing hazards during on-chip data movement.

Consider the routing scenario illustrated in Fig.~\ref{fig:fig1-preliminary}(b) and (c). To execute the subsequent gate $g_{i+1}$, the scheduler must transport atoms $q_2$ and $q_3$ to their target SLM coordinates using AOD rows $r_0$ and $r_1$, respectively. Although the operations might lack logical data dependencies, simultaneously actuating $r_0$ and $r_1$ along intersecting trajectories would violate the non-crossing hardware rule, precipitating a physical collision. To resolve this microarchitectural conflict, the compiler is forced to serialize the routing process. As depicted, the data movement is decomposed into discrete execution stages---relocating $q_2$ in Cycle 1 while deferring the transit of $q_3$ to Cycle 2. Such forced serialization not only incurs substantial latency penalties but also exacerbates idle decoherence, underscoring the extreme cost of kinematic routing in neutral atom processors.

Ultimately, the quantum compilation task reduces to a highly constrained microarchitectural scheduling formulation. To ensure continuous circuit execution, the compiler must concurrently satisfy the static ILP boundaries and the non-crossing data movement rules. However, the true complexity of this formulation stems from the deep \textbf{spatiotemporal entanglement} of these two dimensions. The geometric placement of atoms and the time-evolving dataflow of the quantum circuit are inextricably linked: a spatial configuration that achieves optimal gate density for the current operations frequently dictates congested, or even severely impeded, AOD routing trajectories in subsequent layers. This severe collision between rigid geometric interference and continuous topological evolution creates an intractable search space, establishing the scalability wall for current neutral atom compilers.

\subsection{The Bottleneck of Spatiotemporal Coupling in Related work}

To navigate the rigid microarchitectural constraints established above, compilation methodologies have continuously evolved. Exact solvers \cite{Tan2024, Tan2022, Baker2021} and advanced stochastic frameworks, such as Enola \cite{Tan2025}, provide immense value by achieving highly optimized layouts for small-scale quantum circuits. These pioneering approaches typically adopt a unified formulation, attempting to synthesize the spatial ILP boundaries and the dynamic kinematic rules within a single global constraint satisfaction or annealing process. While mathematically elegant, this joint-optimization paradigm demands that every proposed physical atom placement must simultaneously guarantee a collision-free future AOD routing trajectory. Because the geometric interaction radius and the continuous topological dataflow are non-linearly entangled, any local adjustment to satisfy a spatial constraint inherently forces the compiler to re-evaluate the entire downstream routing graph. Consequently, rather than suffering from algorithmic inefficiency, these methods fall victim to a fundamental combinatorial explosion. The compiler exhausts the vast majority of its execution cycles repeatedly verifying geometric legality across an astronomical number of routing permutations, ultimately hitting a computational wall that severely precludes practical scalability.

To circumvent this computational wall, alternative heuristic frameworks, such as DasAtom \cite{Huang2025}, have adopted a spatial-first scheduling policy driven by exact graph isomorphism algorithms. While this methodology excels at rapidly resolving static spatial boundaries to ensure fast compilation for small-scale circuits, it operates with a critical deficit of topological foresight. By evaluating spatial placements in isolation—without integrating deep temporal routing information or incorporating subsequent gate dependencies—these policies fundamentally fail to sustain high ILP across sequential layers. Consequently, this myopic placement forces the compiler into excessive gate serialization, drastically inflating the total number of error-prone Rydberg excitation stages and precipitating a severe degradation in overall physical execution fidelity. Furthermore, the intrinsic reliance on exact graph matching algorithms introduces a profound structural scalability bottleneck, rendering the framework computationally intractable when navigating the dense geometric complexities of massive qubit arrays.

The severe consequences of these two opposing scheduling traps are empirically substantiated by the scaling trends presented earlier in Fig.~\ref{fig:intro}. As the evaluation indicates, state-of-the-art compilers inevitably fall into a structural dilemma: they either succumb to the exponential time complexity of joint optimization or suffer the fidelity collapse of myopic spatial greediness. The fundamental flaw uniting these disparate methodologies is their shared reliance on compile-time spatiotemporal coupling. Whether attempting to solve both dimensions simultaneously or sacrificing the future to optimize the present, the requirement to evaluate rigid static geometric rules alongside continuous dynamic topologies within the same compilation cycle proves unscalable. 

\subsection{Architectural Insights for ARGON }

Breaking this scheduling bottleneck requires abandoning coupled evaluation. We observe that the spatial and temporal dimensions constraining the compiler are fundamentally heterogeneous in their origin and behavior: static spatial constraints are strictly governed by invariant hardware physics, whereas dynamic kinematic constraints are intrinsically tied to the algorithmic dataflow. This fundamental dichotomy directly yields the two microarchitectural insights that drive ARGON as follow:
\\
\rule{\linewidth}{0.4pt}
\noindent\textbf{\textit{Insight 1:}}
\textbf{Geometric interference rules are hardware - intrinsic and algorithm - invariant, enabling the extraction of spatial constraint resolution from the critical compilation path.}
\rule{\linewidth}{0.4pt}

Driven by this observation, we recognize that because parameters such as the interaction and restriction radii are fixed physical properties of the optical trapping hardware, the theoretical set of valid spatial placements is entirely independent of the specific quantum circuit being executed. Therefore, ARGON strictly decouples spatial ILP resolution by offloading it to an offline phase. By precomputing a discrete library of hardware-certified, maximum-density spatial layouts, the compiler completely eradicates the overhead of geometric verification. Furthermore, this offline offloading mathematically guarantees peak gate parallelism, which directly minimizes the required number of error-prone Rydberg excitation stages and secures the first pillar of physical execution fidelity.
\\
\rule{\linewidth}{0.4pt}
\noindent\textbf{\textit{Insight 2:}}
\textbf{Efficient routing necessitates topological foresight, as myopically prioritizing instantaneous gate parallelism frequently exacerbates global kinematic overhead.}\\
\rule{\linewidth}{0.4pt}

Building upon this realization, we establish that the scheduling phase should transcend greedy spatial matching to operate as a predictive microarchitectural engine. Because AOD movements are strictly bound by non-crossing rules, an atom's current spatial coordinate heavily influences its future routing viability. Therefore, to sustain continuous execution and minimize the injection of decoherence-inducing transport cycles, the compiler must account for the upcoming dataflow dependencies of the quantum circuit. This requisite foresight motivates the integration of a predictive model---specifically, a GNN---into the placement pipeline. By encoding the current physical topology alongside multi-layer future gate dependencies, ARGON evaluates the precomputed layout library to select configurations that not only satisfy immediate spatial constraints but also actively mitigate global downstream routing overhead.

Synthesizing the aforementioned insights, we propose the ARGON framework, a novel compilation paradigm that structurally decouples spatiotemporal constraints during evaluation and subsequently re-integrates them for physical execution. The framework first introduces an offline geometric layout library to completely eradicate the compile-time combinatorial search space. Building upon these hardware-certified layouts, a Graph Neural Network-guided predictive placement engine jointly captures circuit dependencies and physical mapping features to select a suitable layout that minimizes downstream routing congestion.  Finally, a rapid heuristic parallel router re-couples these placement decisions into coordinated, collision-free AOD kinematics. Through this cohesive pipeline, ARGON effectively reconciles local geometric constraint satisfaction with global execution efficiency. The detailed microarchitectural design and workflow of this framework are presented in the next section.

\section{The ARGON Framework}

\begin{figure*}[htbp]
	\centering
	\includegraphics[width=1\textwidth, trim=8cm 10.5cm 8cm 10.5cm, clip]{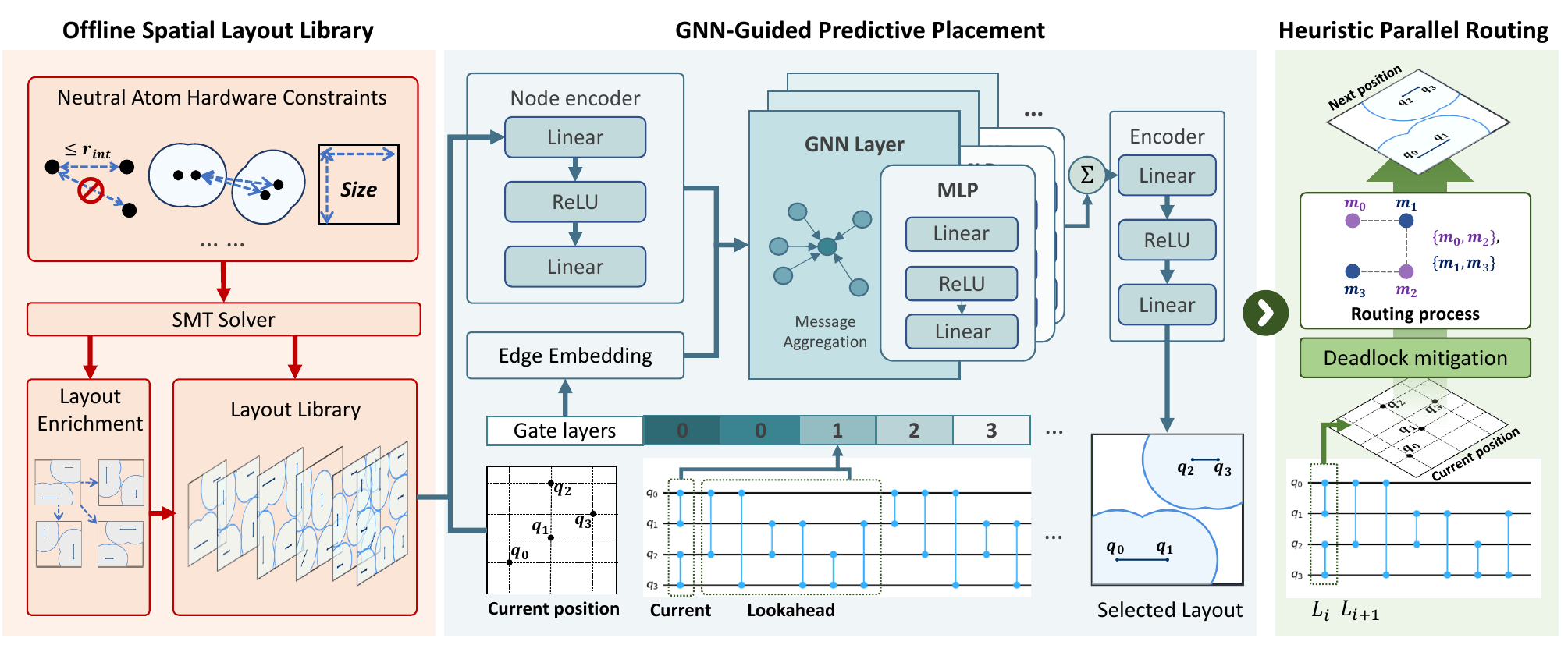}
	\caption{Overview of the ARGON Framework. The compilation pipeline isolates spatiotemporal evaluations into three sequential phases: an offline spatial layout library construction, a GNN-guided predictive temporal placement, and a lightweight heuristic backend router. This decoupling effectively translates hardware-agnostic quantum circuits into collision-free physical transport trajectories.}
	\label{fig:overvierw}
\end{figure*}

To overcome the intractable spatiotemporal coupling, ARGON is architected as a decoupled compilation pipeline illustrated in Fig.~\ref{fig:overvierw}. The framework comprises three integrated modules: First, the \textbf{Offline Spatial Layout Library} (Section~\ref{subsec:libconstruct}) amortizes the computational overhead of geometric verification by precomputing a repository of hardware-certified, maximum-parallelism physical placements. During the compilation process, the \textbf{GNN-Guided Predictive Placement} engine (Section~\ref{subsec:gnn_placement}) evaluates these discrete candidate layouts against deep temporal circuit dependencies, proactively bypassing downstream execution blockages. Finally, the \textbf{Heuristic Parallel Router} (Section~\ref{subsec:routing}) synthesizes these layout decisions into coordinated, collision-free AOD transport trajectories. Through this structural separation, ARGON achieves rapid compilation responsiveness while preserving execution fidelity.

\subsection{Offline Spatial Layout Library Construction}\label{subsec:libconstruct}

Because hardware geometric properties---such as Rydberg interaction and restriction radii---are invariant to the algorithmic logic of the executed circuit, evaluating them during compilation introduces redundant computational overhead. ARGON strictly decouples this by offloading the spatial constraint satisfaction to an offline Non-Recurring Engineering (NRE) phase. By utilizing a Satisfiability Modulo Theories (SMT) solver, we precompute a discrete library of valid spatial layouts that natively guarantee maximum gate parallelism while strictly adhering to physical hardware limits~\cite{Tan2020,Lin2023,Shaik2024}.

To formulate this offline synthesis, we model the neutral atom processor as an $n \times n$ static grid $V \subset \mathbb{Z}^2$. The compilation must satisfy two physical microarchitectural boundaries. First, the proximity constraint bounds two-qubit operations within a maximum interaction radius ($R_{\text{int}}$). Second, the parallel execution constraint mandates that simultaneous gates maintain disjoint exclusion zones defined by a restriction radius ($R_{\text{r}}$) to prevent crosstalk. 

Rather than explicitly navigating this continuous spatial parameter space at compile-time, we map the geometric interference to a Maximum Independent Set (MIS) problem. We define the set of all physically executable gate pairings as $\mathcal{G}_{\text{valid}} = \{ (u, v) \in V \times V \mid 0 < \|u - v\|_2 \leq R_{\text{int}} \}$. To prevent structural hazards, we construct a conflict graph where an edge exists between any two gates $g_i, g_j \in \mathcal{G}_{\text{valid}}$ if their spatial coordinates violate the restriction radius (i.e., $\min_{a \in g_i, b \in g_j} \|a - b\|_2 < R_{\text{r}}$). Introducing a binary decision variable $x_i \in \{0,1\}$ to indicate the activation of gate $g_i$, the solver identifies the maximum-capacity layout by optimizing:
\begin{equation}
	\max \sum_{i=1}^{|\mathcal{G}_{\text{valid}}|} x_i \quad \text{s.t.} \quad x_i + x_j \leq 1, \quad \forall (g_i, g_j) \text{ in conflict}.
	\label{eq:max_sat}
\end{equation}

\begin{figure}[htbp]
	\centering
	\begin{subfigure}{0.48\columnwidth}
		\centering
		\includegraphics[width=\linewidth]{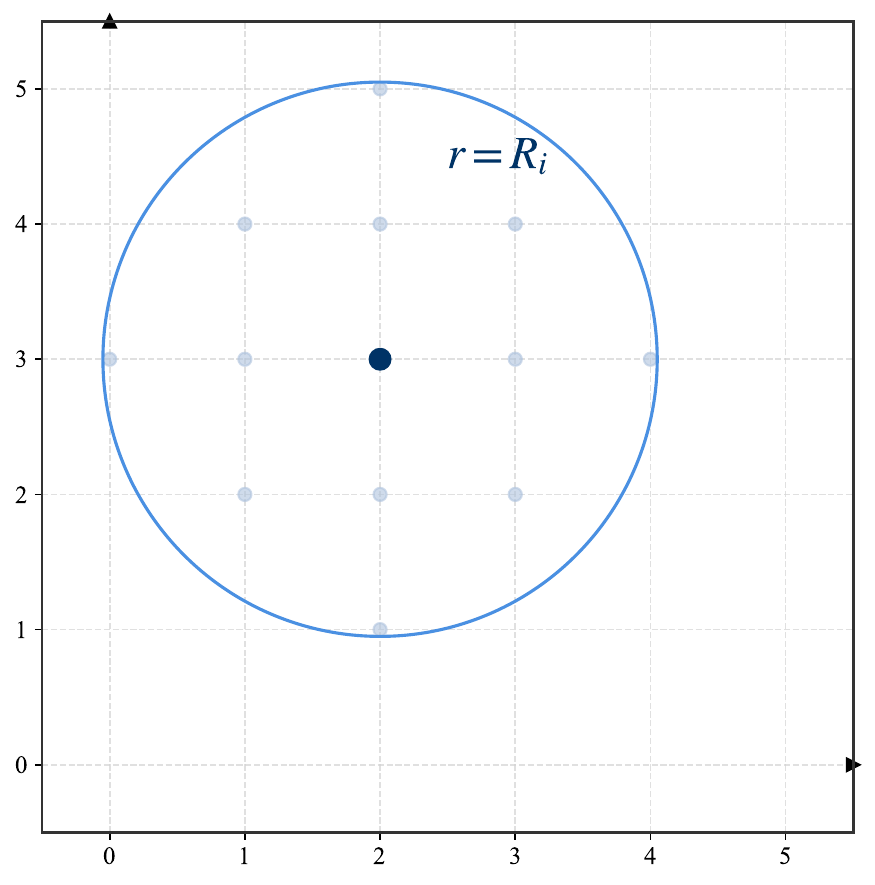}
		\caption{Valid gate set $\mathcal{G}_{\text{valid}}$.}
		\label{fig:rint}
	\end{subfigure}
	\hfill
	\begin{subfigure}{0.48\columnwidth}
		\centering
		\includegraphics[width=\linewidth]{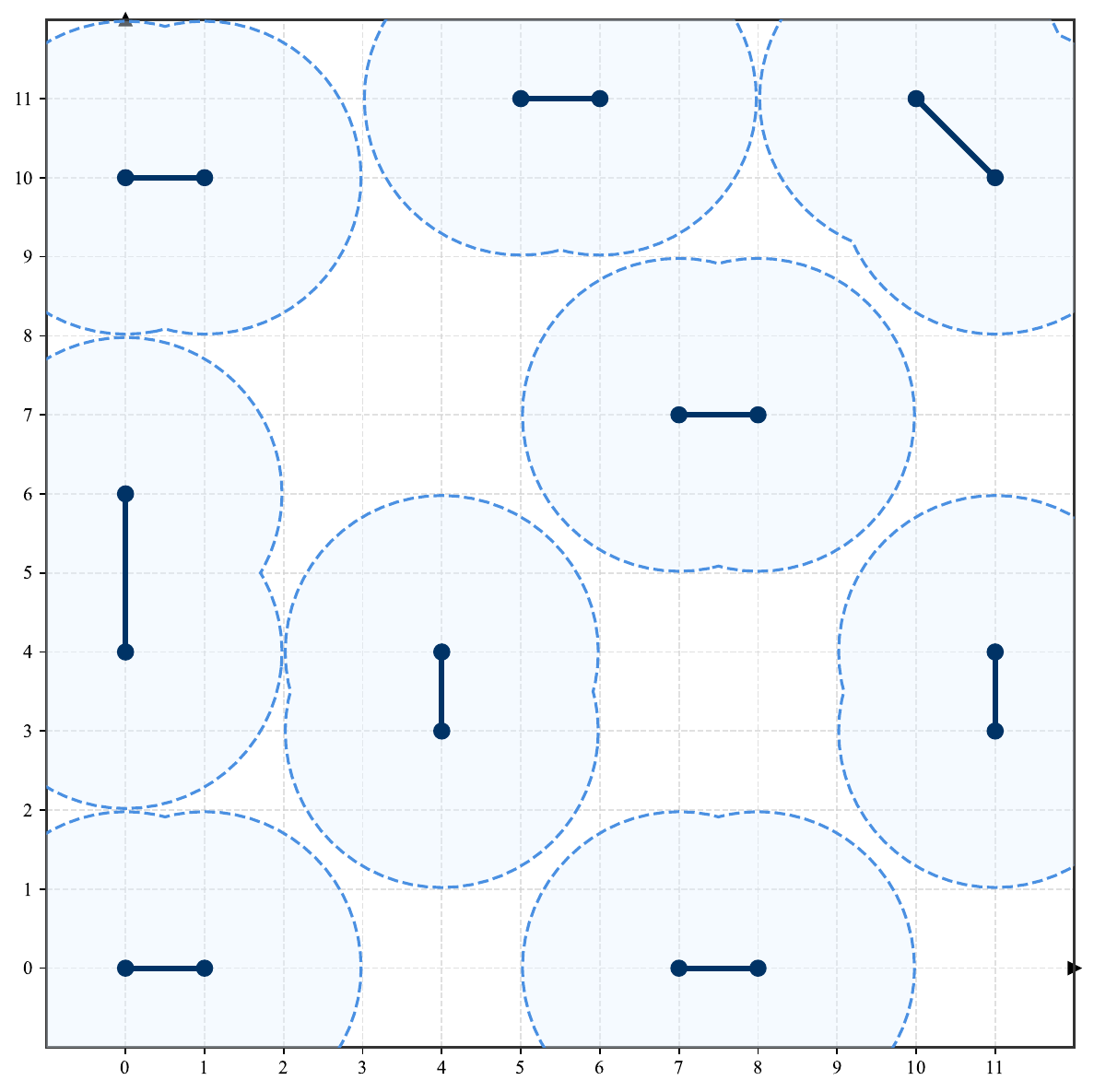}
		\caption{Maximal subset $\mathcal{G}_{\text{select}}$.}
		\label{fig:parallel_execution_map}
	\end{subfigure}
	\caption{Visualization of spatial connectivity and solver optimized parallelism on a neutral atom grid. a Valid connectivity scope: Illustration of all physically feasible two qubit gate pairings for a central atom, constrained by an interaction radius $R_{\text{int}} = 2$ normalized to trap spacing. b Maximal parallel layout: A solver generated layout on a $12 \times 12$ grid that maximizes the number of simultaneous gates $\mathcal{G}_{\text{select}}$. The solution achieves dense spatial packing while enforcing disjoint restriction zones to prevent parallel execution violations.}
	\label{fig:precompile-solver}
\end{figure}

As detailed in Algorithm~\ref{alg:gate_selection}, the solver iteratively sweeps across varying grid dimensions and target gate densities to extract the maximal subset $\mathcal{G}_{\text{select}}$. As depicted in Fig.~\ref{fig:precompile-solver}, this generates highly packed spatial configurations with geometrically disjoint restriction zones. These foundational layouts are subsequently expanded via orthogonal rotations and reflections to maximize topological diversity with near-zero computational cost. 

From a system architecture perspective, the core advantage of this offline methodology is twofold. First, it mathematically bounds the required Rydberg excitation stages by maximizing static ILP, thereby securing baseline physical fidelity~\cite{Tan2025}. Second, it completely eradicates the compile-time combinatorial search space, providing a highly compressed, hardware-certified discrete decision library for the subsequent predictive routing phase.

\begin{algorithm}[t]
	\caption{Offline Spatial Layout Generation}
	\label{alg:gate_selection}
	\KwIn{Grid dimension $n$, Interaction radius $R_{\text{int}}$, Restriction Radius $R_{\text{r}}$}
	\KwOut{Selected Layout $\mathcal{G}_{\text{select}}$}
	$\mathcal{G}_{\text{valid}} \leftarrow \emptyset$;\\
	\ForEach{pair $u, v$ in $n \times n$ grid}{
		\If{$\|u - v\|_2 \leq R_{\text{int}}$ and $u < v$} {
			Add $(u, v)$ to $\mathcal{G}_{\text{valid}}$;
		}
	}
	Construct conflict graph $\mathcal{C}$: nodes $\mathcal{G}_{\text{valid}}$, edges for overlapping restriction zones;\\
	Formulate Max SAT constraints for Independent Set on $\mathcal{C}$;\\
	$\mathcal{G}_{\text{select}} \leftarrow \text{SMT\_Solver}(\text{Constraints}, \text{Objective}=\max \sum x_i)$;\\
	\Return{$\mathcal{G}_{\text{select}}$}
\end{algorithm}

\subsection{GNN-Guided Predictive Placement}\label{subsec:gnn_placement}

\subsubsection{Microarchitectural Feature Encoding and Predictive Policy}

\begin{figure}[htbp]
	\centering
	\includegraphics[width=0.6\textwidth, trim=16.5cm 12.5cm 12.5cm 12.5cm, clip]{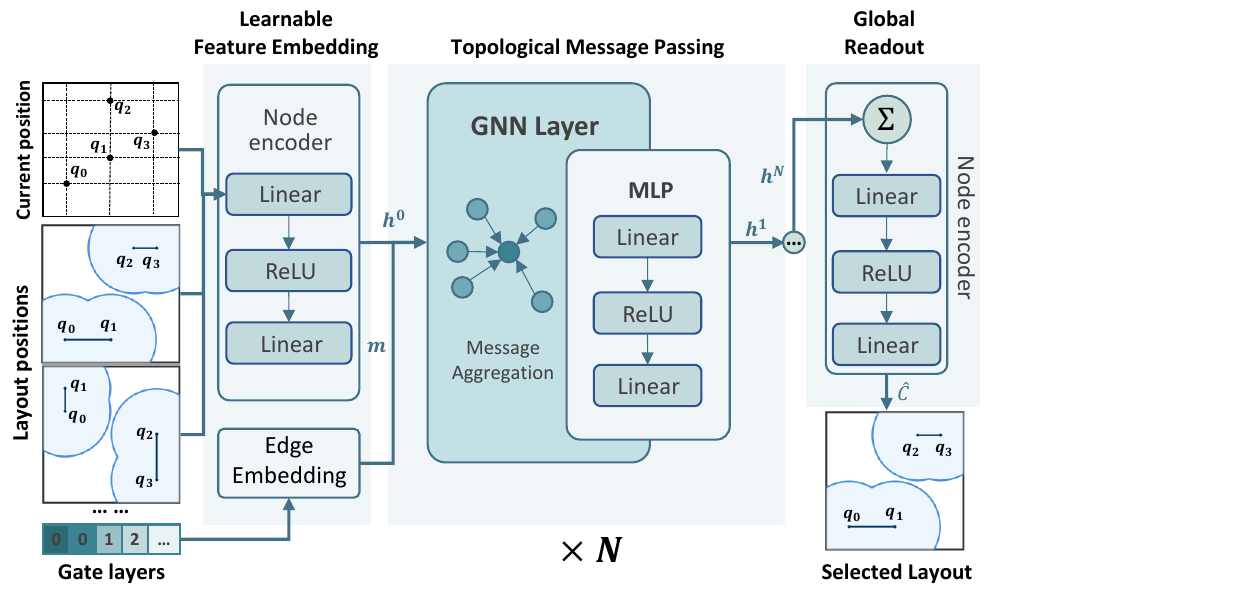}
	\caption{The architectural structure of the GNN-guided predictive placement engine.}
	\label{fig:nn_structure}
\end{figure}

While the offline library resolves static geometric constraints, the compiler must select a layout that minimizes temporal routing overhead. Myopic layout selection frequently traps qubits in physical coordinates that precipitate AOD row or column crossing conflicts in subsequent layers. To navigate this, ARGON introduces a predictive graph neural policy. 

We adopt a GNN rather than traditional search heuristics for two fundamental architectural reasons. First, GNNs natively encode both the physical geometry of the atom array and the evolving logical topology, enabling the structural awareness required for long-horizon routing predictions. Second, this end-to-end model fundamentally transforms the system overhead distribution. Evaluating deep routing topologies is inherently high-overhead; however, the neural paradigm absorbs this computational burden entirely during the offline dataset construction phase. Consequently, layout evaluation is reduced to a constant-time neural forward pass, allowing the compiler to strictly satisfy practical compilation latency constraints.

To operationalize this prediction (Fig.~\ref{fig:nn_structure}), the compiler maps the current layer's logical gates to the candidate layout's physical slots via a lightweight distance-minimization heuristic. For each active qubit, we extract a 4D kinematic descriptor explicitly encoding the required physical AOD actuation trajectory: $[x_{\text{old}}, y_{\text{old}}, x_{\text{new}}, y_{\text{new}}]$. To provide topological foresight, future gate dependencies within a lookahead window $K$ are extracted as graph edges, each assigned a discrete time-lag attribute $\tau$.

\begin{figure}[htbp]
	\centering
	\includegraphics[width=0.7\linewidth]{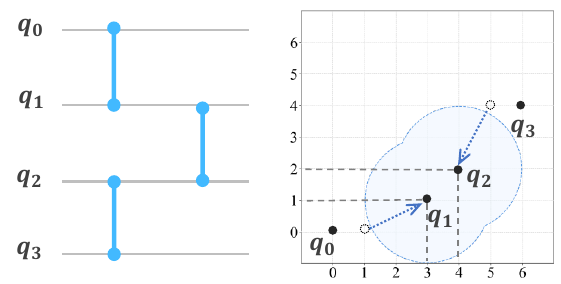}
	\caption{An example of encoding $\mathbf{X}$, $\tau$, and $\mathbf{E}$}
	\label{fig:encoding-example}
\end{figure}

Fig.~\ref{fig:encoding-example} illustrates this encoding for four qubits $q_0$–$q_3$. Layer $t$ contains gates $(q_0,q_1)$ and $(q_2,q_3)$, followed by $(q_1,q_2)$ in layer $t+1$. Under a candidate layout, $q_1$ and $q_2$ are reassigned to $(3,1)$ and $(4,2)$ to execute gate $(q_1,q_2)$, respectively. The node feature matrix $\mathbf{X}$, temporal vector $\tau$, and edge index tensor $\mathbf{E}$ are then encoded as:

\begin{equation}
	\mathbf{X} = 
	\begin{bmatrix} 
		0 & 0 & 0 & 0 \\ 
		1 & 0 & 3 & 1 \\ 
		5 & 4 & 4 & 2 \\
		6 & 4 & 6 & 4 
	\end{bmatrix}, \quad
	\tau = \begin{bmatrix} 0 & 0 & 1 \end{bmatrix}, \quad
	\mathbf{E} = 
	\begin{bmatrix} 
		0 & 2 & 1\\ 
		1 & 3 & 2
	\end{bmatrix}.
\end{equation}
This contextual encoding allows the predictive policy to foresee impending trajectory collisions and select an alternative layout for layer $t$ that maintains global AOD routability.

To process these inputs, the node feature matrix $\mathbf{X}$ is linearly projected to form the initial hidden state $\mathbf{h}_i^{(0)}$ for each qubit $i$. Concurrently, the temporal vector $\tau$ and edge tensor $\mathbf{E}$ are jointly embedded to construct $\mathbf{m}_{ji}$, which models the physical spatial transition from neighboring node $j$ to the central node $i$. These representations are processed through three layers of an extended Graph Isomorphism Network. At layer $l$, the hidden state of each qubit $i$ is updated to aggregate localized AOD movement penalties across the array:

\begin{equation}
	\mathbf{h}_i^{(l)} = \text{MLP}^{(l)} \left( (1+\epsilon)\mathbf{h}_i^{(l-1)} + \sum_{j \in \mathcal{N}(i)} \text{ReLU}(\mathbf{h}_j^{(l-1)} + \mathbf{m}_{ji}) \right),
\end{equation}
where $\epsilon$ is a learnable scalar to preserve representation injectivity.

Finally, a global sum-pooling operation coalesces these distributed hidden states into a unified compilation cost prediction $\hat{C}$:

\begin{equation}
	\hat{C} = \text{MLP}_{\text{head}} \left( \sum_{i \in V} \mathbf{h}_i^{(L)} \right).
\end{equation}
We deliberately adopt sum-pooling because microarchitectural routing penalties---such as cumulative atom transit distances and localized trajectory delays---are inherently additive across the processor.

\subsubsection{Offline Rollout Training and Compile-Time Inference Overhead}

As established by recent literature \cite{Wang2024, Tan2025, Ruan2025}, acquiring exact optimal routing-cost evaluations is inherently NP-hard. To parameterize the predictive policy, we deliberately embrace this massive computational complexity during the offline dataset construction phase. This deliberate architectural asymmetry---amortizing computationally intensive spatiotemporal rollout evaluations into offline neural weight optimization---establishes the foundation for ARGON's scalable responsiveness.

To generate high-fidelity supervisory signals without exhaustive search, we employ a computationally intensive spatiotemporal rollout strategy. For a given circuit layer $L_t$ and a candidate layout, the simulator instantiates the physical state $S_t$ and computes the immediate transition cost $c_{\text{local}}(S_t)$, which explicitly penalizes maximum AOD transit distances and movement violations. The simulator then executes a deterministic heuristic rollout across a lookahead window of $K$ layers. At each step $k$, the simulator systematically compares the routing performance of a vast space of valid subsequent layout templates. It then selects the one minimizing the immediate routing overhead, yielding a highly optimized simulated trajectory. The target cumulative cost $C$ is formulated as the discounted sum of these penalties:
\begin{equation}
	C = c_{\text{local}}(S_t) + \sum_{k=1}^{K} \gamma^k \cdot c_{\text{local}}(S_{t+k}),
\end{equation}
where $\gamma \in (0, 1]$ attenuates the influence of distant future operations, reflecting the diminishing certainty of long-term predictions.

Executing this rollout across diverse benchmark circuits yields a comprehensive dataset $\mathcal{D} = \{(\mathbf{G}_m, C_m)\}_{m=1}^M$, directly linking the spatiotemporal graph representations to their global routing values. The network parameters $\mathbf{\Theta}$ are optimized via supervised learning to minimize the Mean Squared Error (MSE):
\begin{equation}
	\mathcal{L}(\mathbf{\Theta}) = \frac{1}{M} \sum_{m=1}^{M} \left( \hat{C}_m(\mathbf{\Theta}) - C_m \right)^2.
\end{equation}

Crucially, this supervised learning paradigm acts as an architectural overhead absorption mechanism. The severe combinatorial complexity of deep lookahead search---which traditionally establishes the scalability wall in compilers attempting joint spatiotemporal optimization---is entirely amortized into the offline training phase. Consequently, during compilation, the exponential exploration of future microarchitectural states is replaced by a constant-time neural forward pass with low cost. This structural shift from compile-time search to offline-trained inference definitively secures ARGON's sub-second responsiveness without sacrificing execution fidelity.

\subsection{Heuristic Parallel Routing}\label{subsec:routing}

The final phase of ARGON serves as the backend instruction generator, translating the layouts selected by the predictive policy into coordinated AOD transport sequences. Since spatial constraints and topological dependencies have been resolved by the preceding modules, routing is reduced from a coupled spatiotemporal optimization problem to a kinematic pathfinding task. The router models atom movements as a directed kinematic graph and synthesizes collision-free AOD actuations along orthogonal axes, ensuring non-intersecting trajectories across rows and columns.

First, the router detects and resolves kinematic deadlocks arising from cyclic target dependencies. When a cycle is encountered, one qubit within the loop is temporarily assigned to an intermediate parking location, converting the cyclic dependency into acyclic movements. This auxiliary step effectively breaks the deadlock while avoiding costly global re-routing and introducing just a small additional movement cost. Safe eviction paths are also generated for idle qubits to prevent interference with active AOD sweeps, thereby guaranteeing collision-free transport.

Second, the resulting acyclic movements are packed into parallel execution slices. We formulate scheduling as a graph coloring problem, where nodes represent AOD trajectories and edges indicate trajectory intersections. A greedy coloring heuristic groups non-conflicting trajectories into the same color class, corresponding to a parallel AOD execution cycle.

By isolating routing into a dedicated backend stage, ARGON enables collision-free execution while efficiently converting decoupled compilation decisions into executable hardware instructions.

\section{Evaluation}
\subsection{Experimental settings}

\textbf{Baselines \& Benchmarks.}
We evaluate ARGON against three recent compilers representing divergent spatiotemporal scheduling philosophies: \textbf{DasAtom}~\cite{Huang2025}, representing the spatial-first heuristic paradigm; \textbf{Enola}~\cite{Tan2025}, representing the joint-optimization approach based on graph edge-coloring; and \textbf{PowerMove}~\cite{Ruan2025}, an optimized heuristic scheduler. For equitable comparisons, PowerMove is configured in its "no-storage" mode to align with monolithic-array assumptions, and Enola operates in its higher-fidelity "dynamic" mode, as its static variant is already surpassed by DasAtom in compilation speed~\cite{Huang2025}.

To comprehensively evaluate routing efficiency and structural scalability, we select a robust set of benchmarks comprising both synthetic microarchitectural stressors and representative real-world quantum applications. The synthetic suites (scaling from $5$ to $60$ qubits) include: (1) \textbf{W-state}~\cite{Quetschlich2023} to evaluate compilation efficiency on strictly structured topologies; (2) \textbf{Random gates}~\cite{Quetschlich2023} to test robustness against unstructured logical dataflows; (3) \textbf{3-Regular Graphs}~\cite{Tan2024} to stress dynamic kinematics under complex long-range connectivities; and (4) \textbf{Quantum Volume (QV)}~\cite{Cross2019}, used to evaluate routing efficiency under high circuit complexity, with circuit scales ranging from 5 to 50 qubits and a depth defined as $2 \times \text{qubits}$. Furthermore, to evaluate ARGON's practical utility on representative workloads, we incorporate (5) \textbf{QASMBench}~\cite{Li2023a}, a comprehensive benchmark suite encompassing diverse practical quantum algorithms and routines. Consistent with established neutral-atom compilation practices~\cite{Tan2024, Huang2025}, all circuits are natively decomposed into the hardware-executable $\{CZ, H, S, T, R_x, R_y, R_z\}$ instruction set.

\textbf{Evaluation Metrics \& Hardware Alignment.}
We assess compilation efficiency via \textbf{Compilation time ($t$)} and physical execution cost via \textbf{Rydberg stages ($L$)}. Minimizing $L$ is microarchitecturally critical, as it directly bounds the error accumulation from both active entangling gates and parasitic idle excitations. To guarantee strict system-level fairness and simulate realistic reconfigurable neutral atom arrays, we evaluate the End-to-End Physical Fidelity ($F$) under a unified global Rydberg pulse constraint. By standard convention, we isolate single-qubit errors to exclusively highlight spatiotemporal routing overheads~\cite{Tan2025,Huang2025}. The overall fidelity is thus formulated as a unified microarchitectural penalty model:

\begin{equation}
	\label{equ:fid_compact}
	F = \underbrace{f_{2}^{|G_2|} \cdot f_{\mathrm{exc}}^{|Q| \cdot L - 2|G_2|}}_{\text{Gate \& Idle Penalty}} \times \underbrace{f_{\mathrm{trans}}^{N_{\mathrm{trans}}}}_{\text{Transfer Penalty}} \times \underbrace{\prod_{q \in Q} \exp \left( -\frac{T_q}{T_2} \right)}_{\text{Decoherence Limit } (F_{\mathrm{dec}})}
\end{equation}

Here, $f_{2}$, $f_{\mathrm{exc}}$, and $f_{\mathrm{trans}}$ denote the two-qubit gate, idle-excitation, and single-atom transfer fidelities, respectively. The variable $T_q$ accounts for the absolute total wall-clock time of each qubit $q$ across the entire schedule, encompassing all idle periods, active transport, and gate execution phases, calculated using the hardware timings ($T_{\mathrm{trans}}$, $T_{\mathrm{1Q}}$, $T_{\mathrm{2Q}}$) and kinematic acceleration $a$ detailed in Table~\ref{tab:hardware}. This formulation accurately models the exponential transverse coherence decay ($F_{\mathrm{dec}}$) over the global execution timeframe.

\begin{table}[ht]
	\centering
	\small
	\caption{Hardware parameters used in the evaluation.}
	\label{tab:hardware}
	\renewcommand{\tabularxcolumn}[1]{m{#1}}
	\begin{tabularx}{\linewidth}{c >{\raggedright\arraybackslash}X c}
		\toprule
		\textbf{Symbol} & \multicolumn{1}{c}{\textbf{Description}} & \textbf{Value} \\
		\midrule
		$f_2$           & Two-qubit gate fidelity                     & 0.995 \\
		$f_{\mathrm{exc}}$  & Idle-qubit fidelity under Rydberg excitation & 0.9975 \\
		$f_{\mathrm{trans}}$ & Single-atom transfer fidelity               & 0.999 \\
		$T_2$           & Qubit coherence time                        & $1.5 \cdot 10^{6}\ \mu s$ \\
		$T_{\mathrm{trans}}$ & Atom transfer time                         & $300\ \mu s$ \\
		$T_{\mathrm{1Q}}$ & Local single-qubit gate time    & $8\ \mu s$ \\
		$T_{\mathrm{2Q}}$ & Two-qubit gate time  & $270\ ns$ \\
		$a$             & Atom movement acceleration                  & $2750\ m/s^2$ \\
		\bottomrule
	\end{tabularx}
\end{table}

\textbf{Hardware Configurations \& Implementation.} To ensure a rigorous microarchitectural comparison, all frameworks are evaluated under identical hardware parameters detailed in Table~\ref{tab:hardware}, calibrated according to state-of-the-art neutral atom processors~\cite{Bluvstein2024, Bluvstein2025}. We adopt a high-density $16 \times 16$ array with atom spacing $d=3\mu \text{m}$, interaction radius $R_{\text{int}} = 6\mu \text{m}$, and restriction radius $R_r = 12\mu \text{m}$, establishing a physically rigorous testbed for reconfigurable architectures. All methods are implemented in Python with PyTorch and Z3-solver and evaluated on a workstation equipped with an AMD Ryzen 9 7900X CPU, 48 GB RAM, and an NVIDIA RTX 4070 Ti SUPER GPU. The rollout training set contains only Random gates and W-state circuits, while Quantum Volume, 3-Regular, and QASMBench are reserved exclusively for held-out evaluation.

\subsection{Compilation System-Level Performance}

\begin{figure*}[t]
\centering
	\begin{subfigure}{0.19\textwidth}
		\centering
		\includegraphics[width=\linewidth]{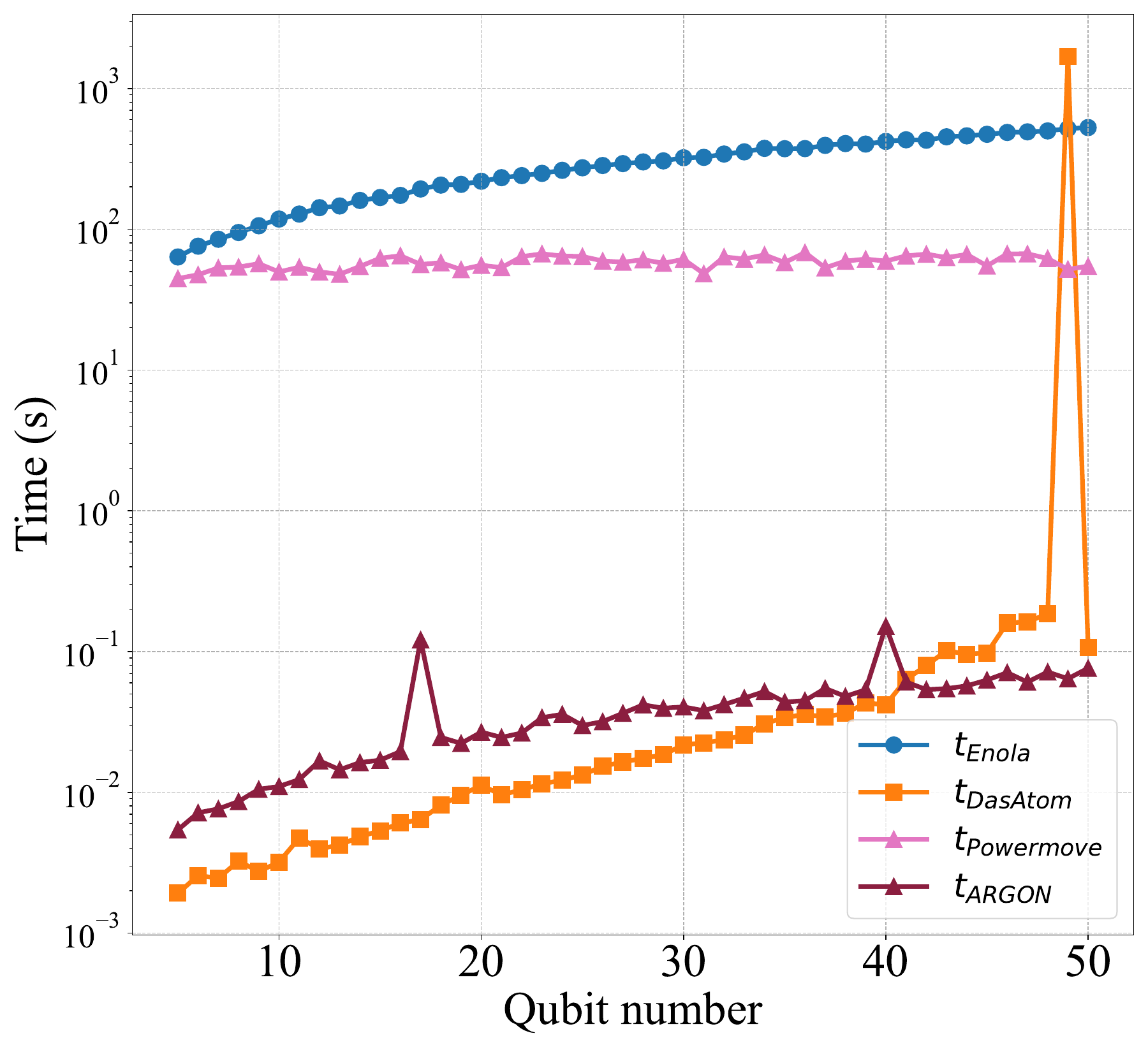}
		\caption{W-state: Time}
	\end{subfigure}
	\hfill
	\begin{subfigure}{0.19\textwidth}
		\centering
		\includegraphics[width=\linewidth]{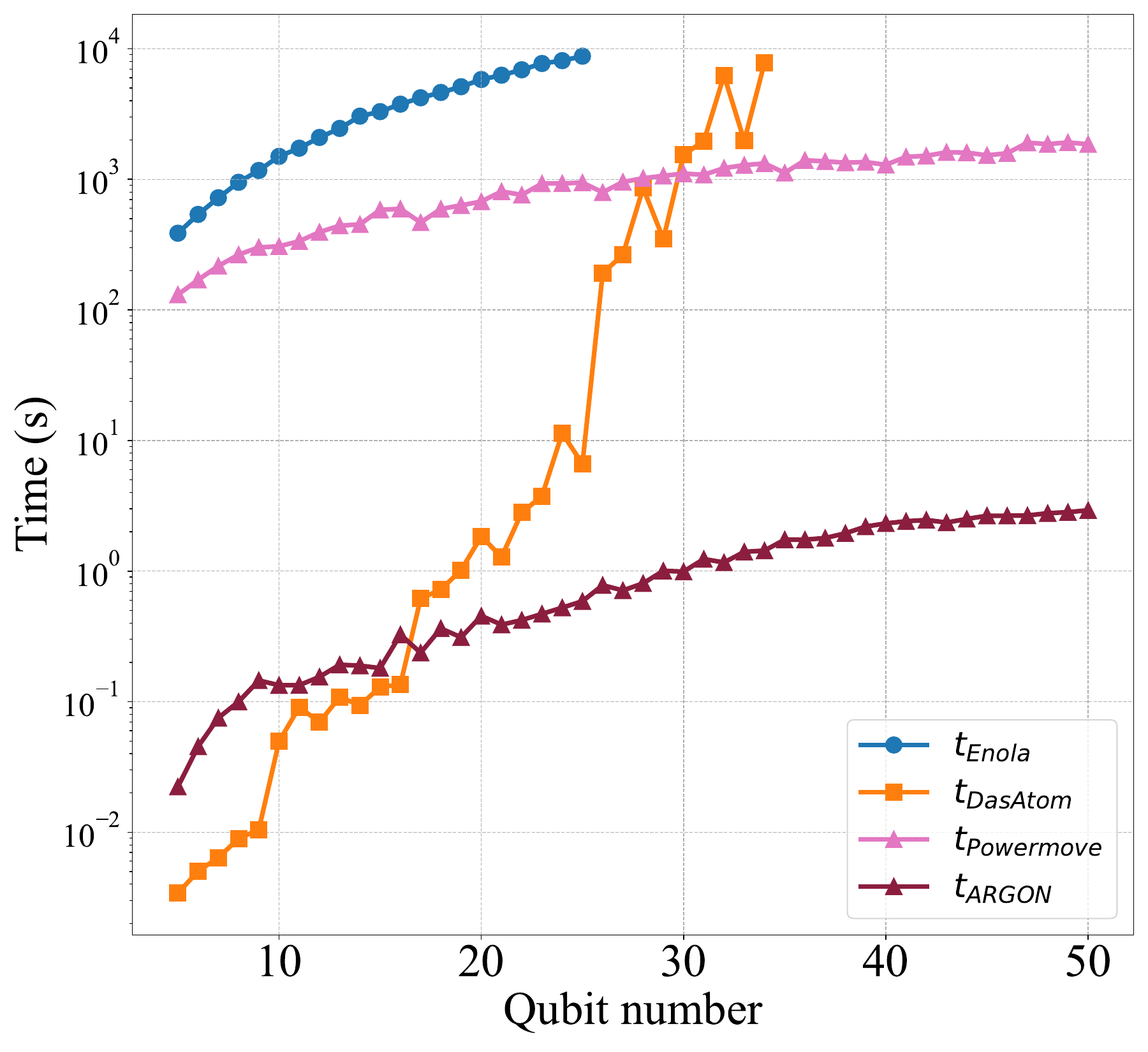}
		\caption{Random gates: Time}
	\end{subfigure}
	\hfill
	\begin{subfigure}{0.19\textwidth}
		\centering
		\includegraphics[width=\linewidth]{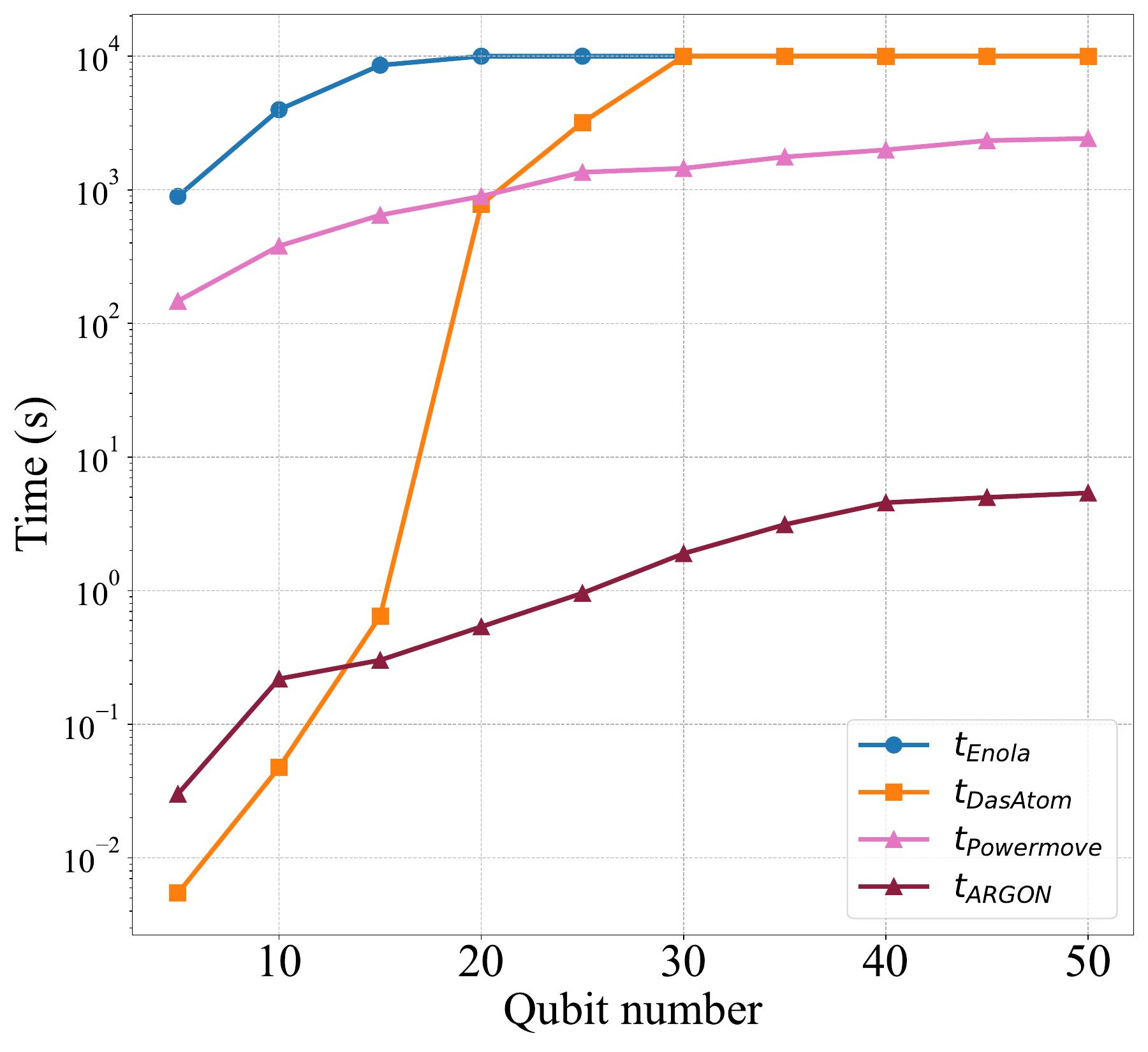}
		\caption{QV: Time}
	\end{subfigure}
	\hfill
	\begin{subfigure}{0.19\textwidth}
		\centering
		\includegraphics[width=\linewidth]{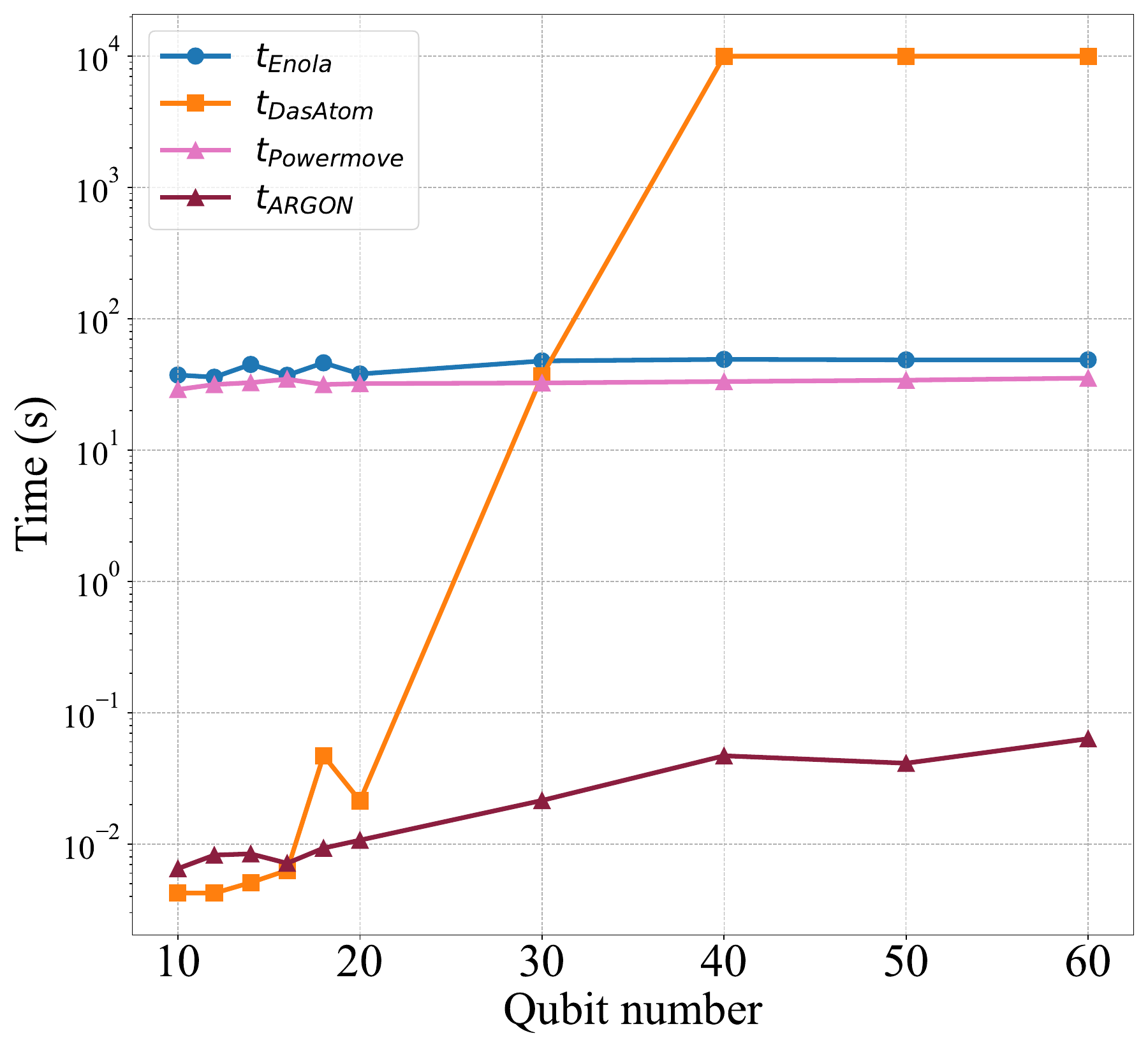}
		\caption{3-Regular: Time}
	\end{subfigure}
	\hfill
	\begin{subfigure}{0.19\textwidth}
		\centering
		\includegraphics[width=\linewidth]{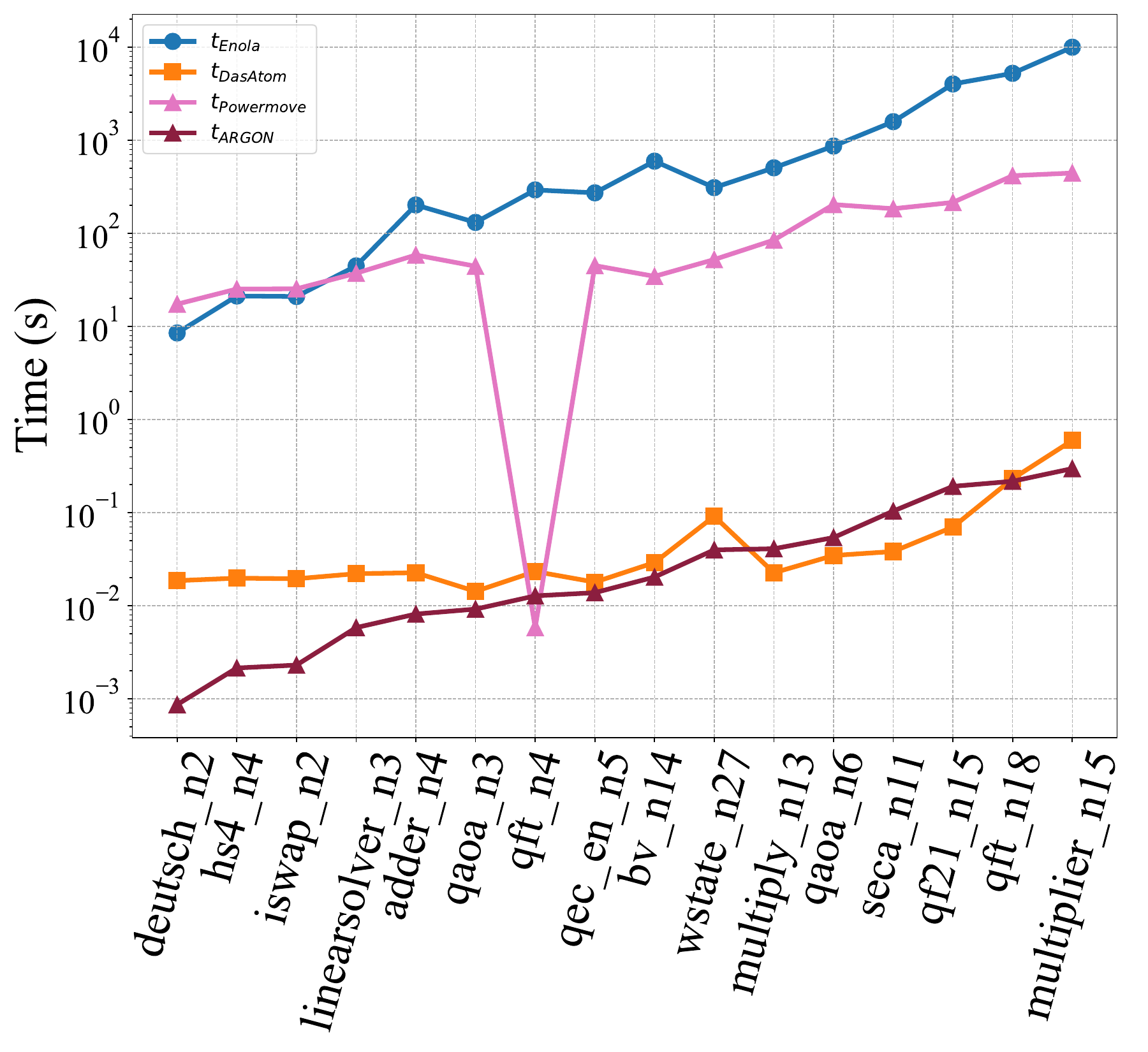}
		\caption{QASMbench: Time}
	\end{subfigure}
	
	% --- 第二行：Fidelity ---
	\begin{subfigure}{0.19\textwidth}
		\centering
		\includegraphics[width=\linewidth]{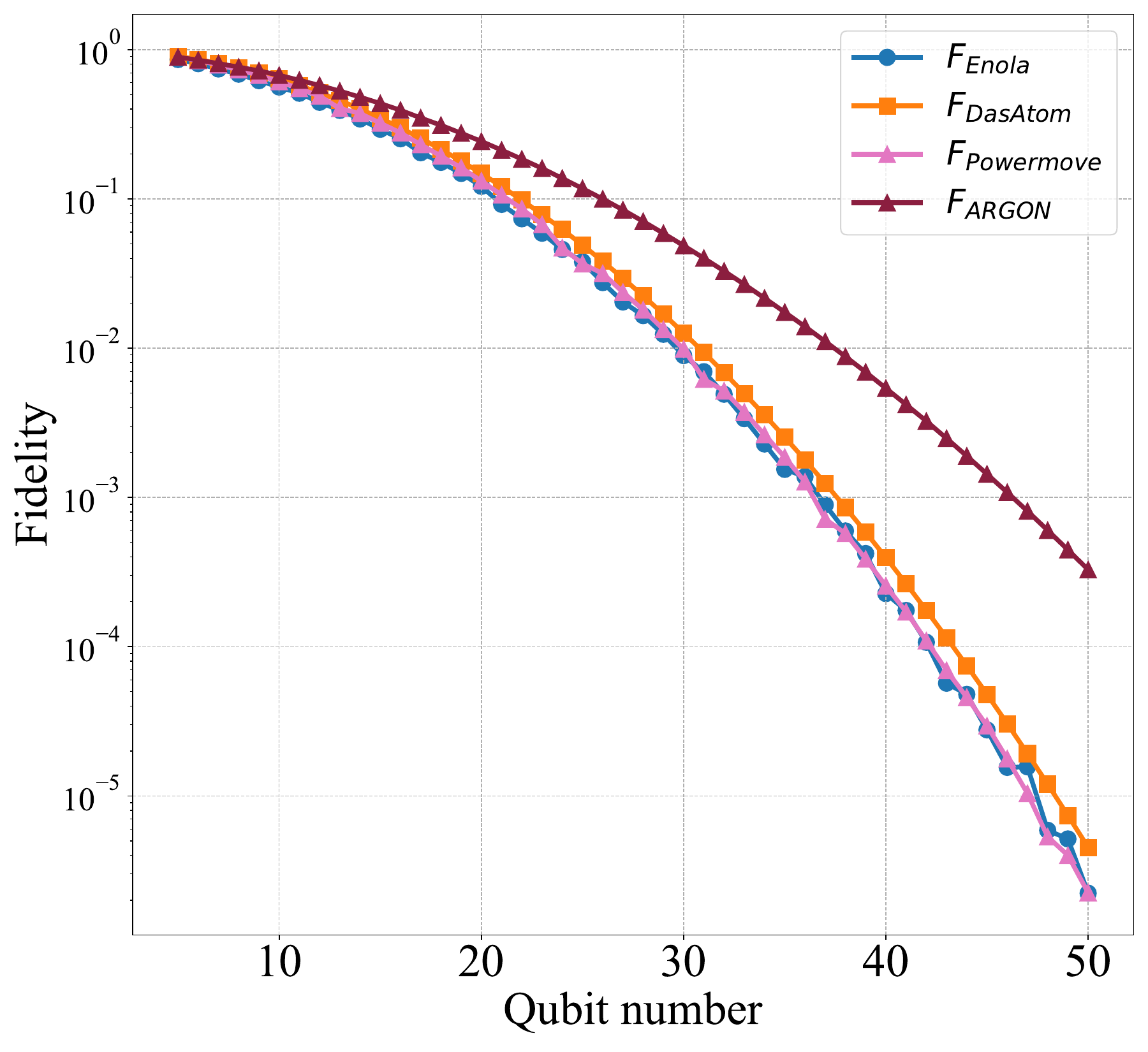}
		\caption{W-state: Fidelity}
	\end{subfigure}
	\hfill
	\begin{subfigure}{0.19\textwidth}
		\centering
		\includegraphics[width=\linewidth]{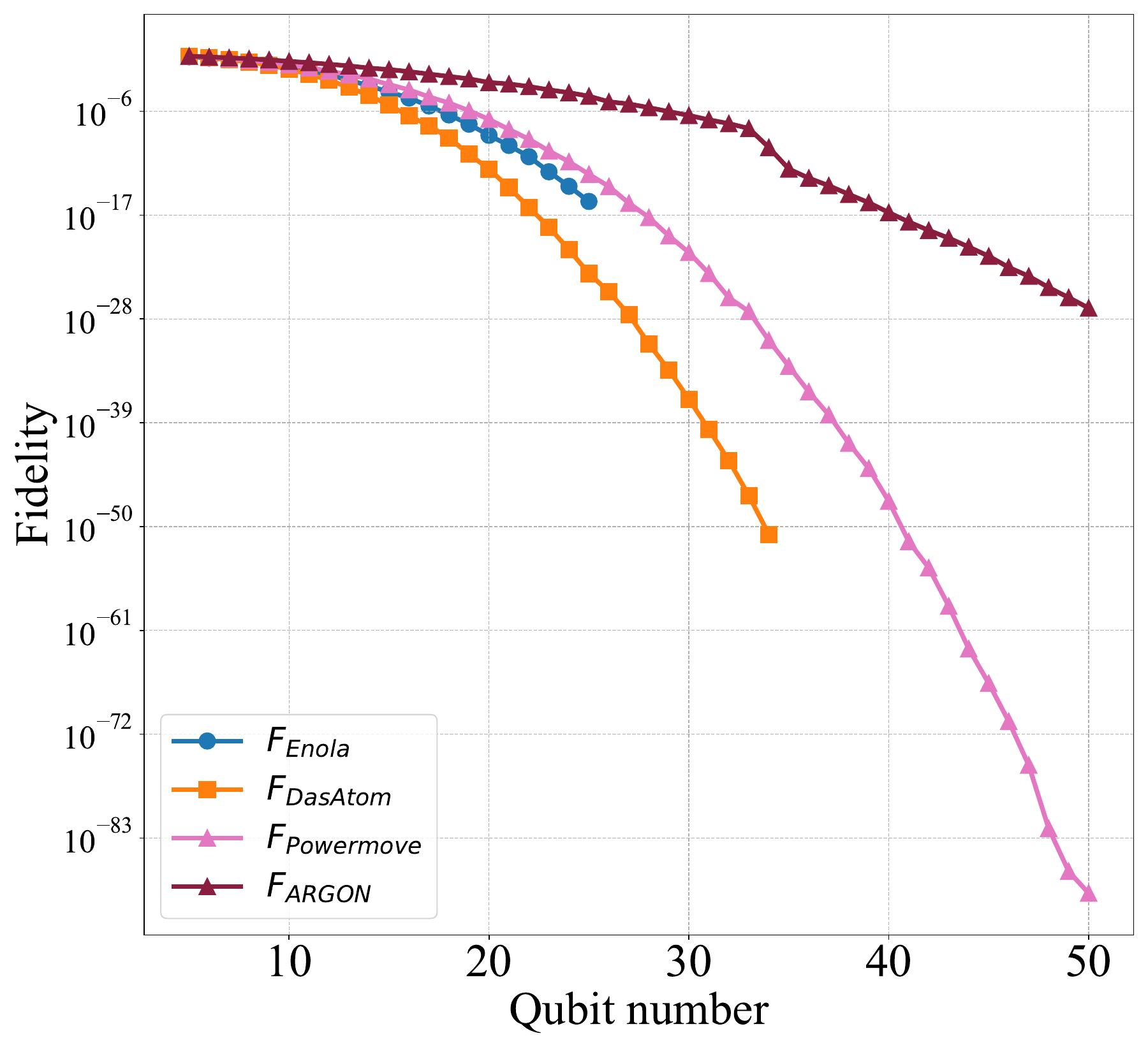}
		\caption{Random gates: Fidelity}
	\end{subfigure}
	\hfill
	\begin{subfigure}{0.19\textwidth}
		\centering
		\includegraphics[width=\linewidth]{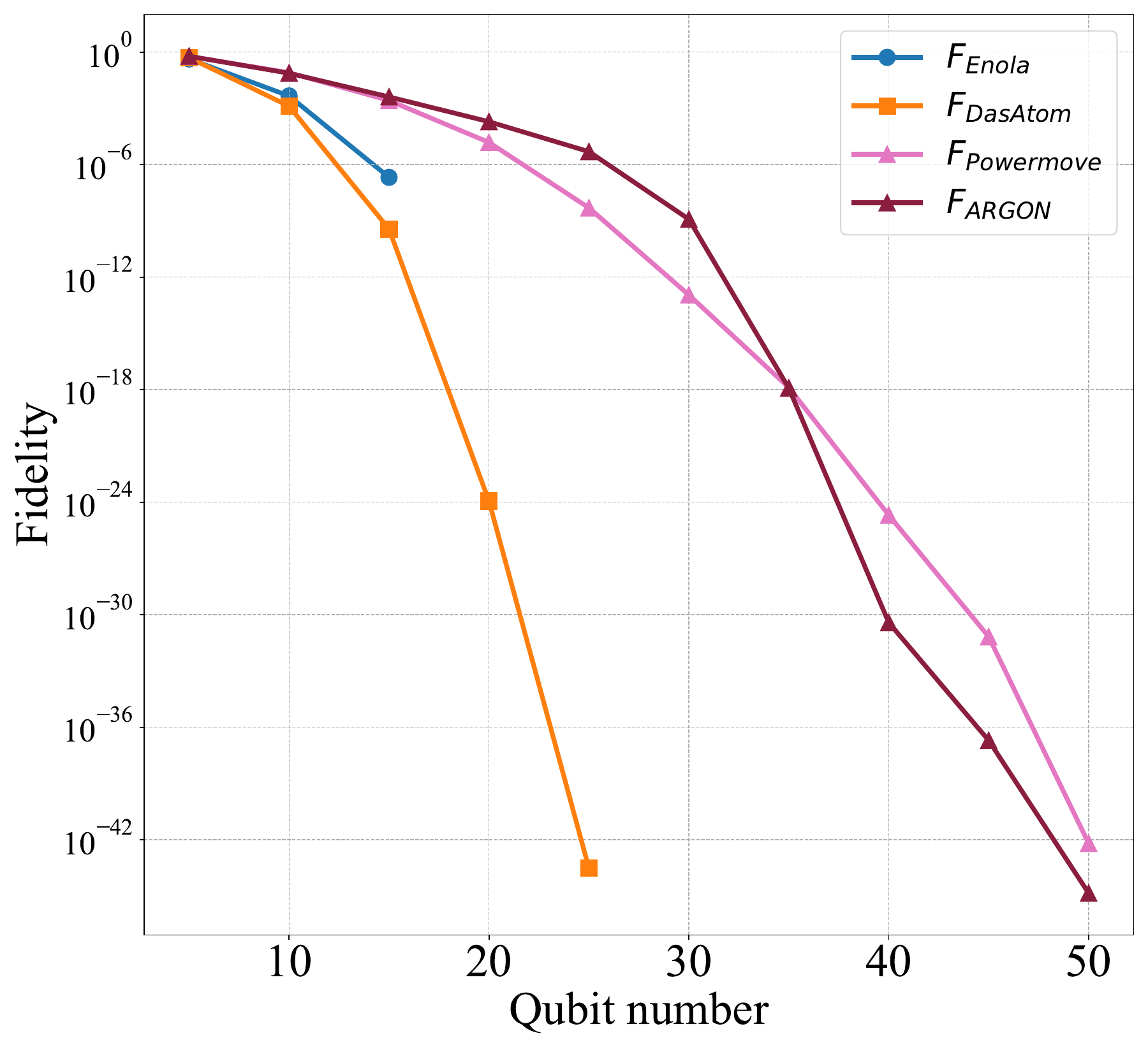}
		\caption{QV: Fidelity}
	\end{subfigure}
	\hfill
	\begin{subfigure}{0.19\textwidth}
		\centering
		\includegraphics[width=\linewidth]{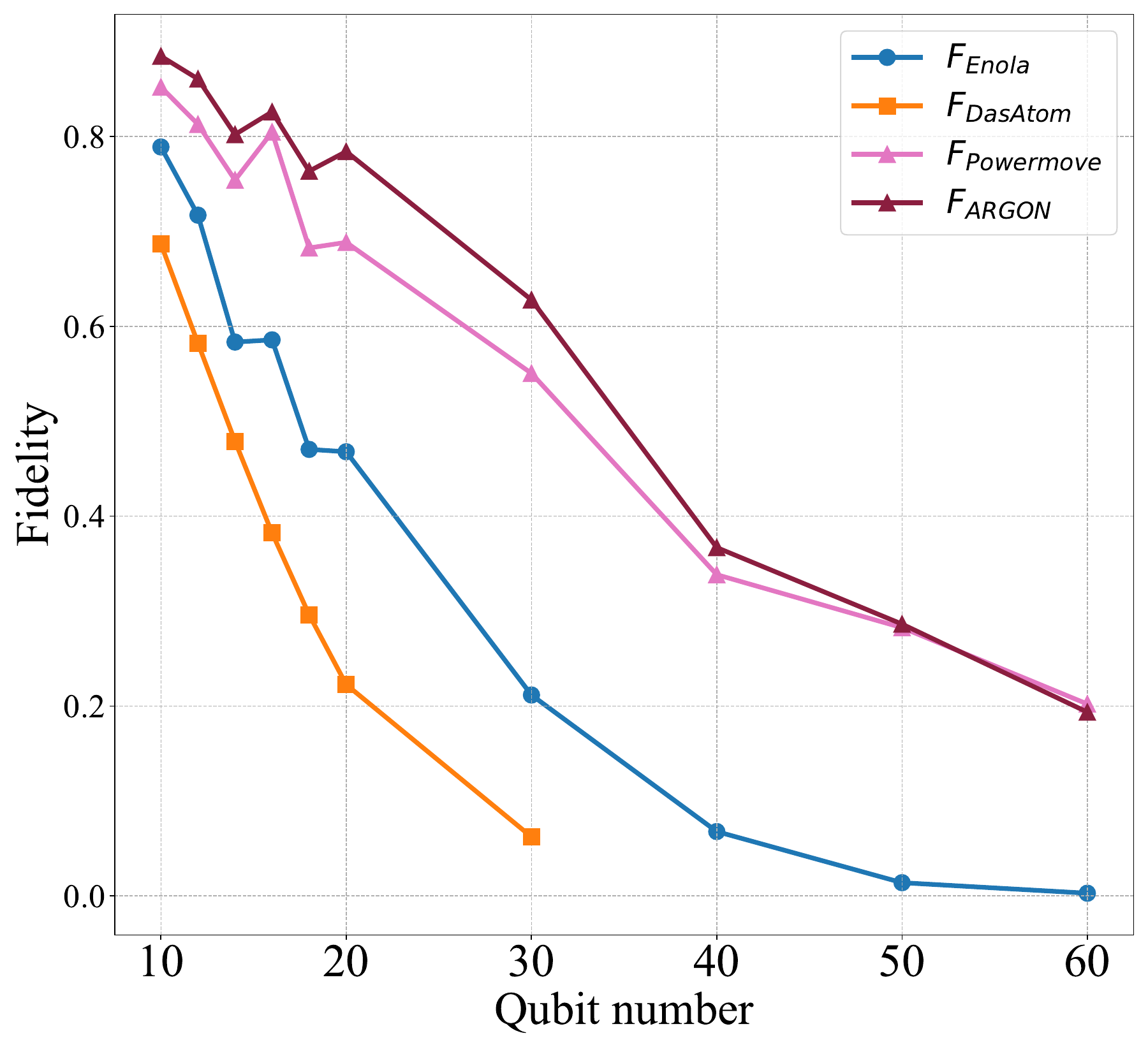}
		\caption{3-Regular: Fidelity}
	\end{subfigure}
	\hfill
	\begin{subfigure}{0.19\textwidth}
		\centering
		\includegraphics[width=\linewidth]{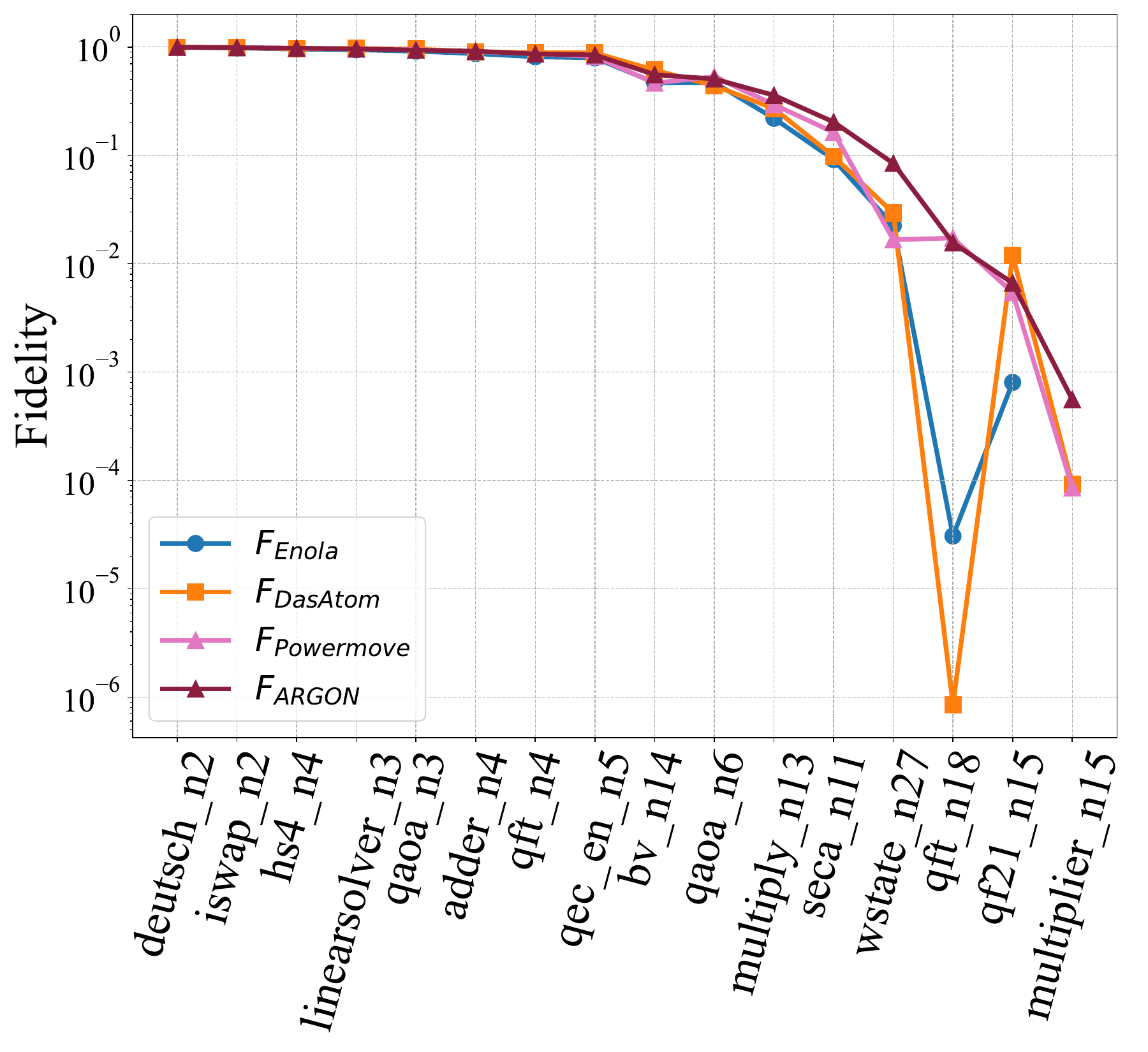}
		\caption{QASMbench: Fidelity}
	\end{subfigure}
	
	% --- 第三行：Rydberg stages ---
	\begin{subfigure}{0.19\textwidth}
		\centering
		\includegraphics[width=\linewidth]{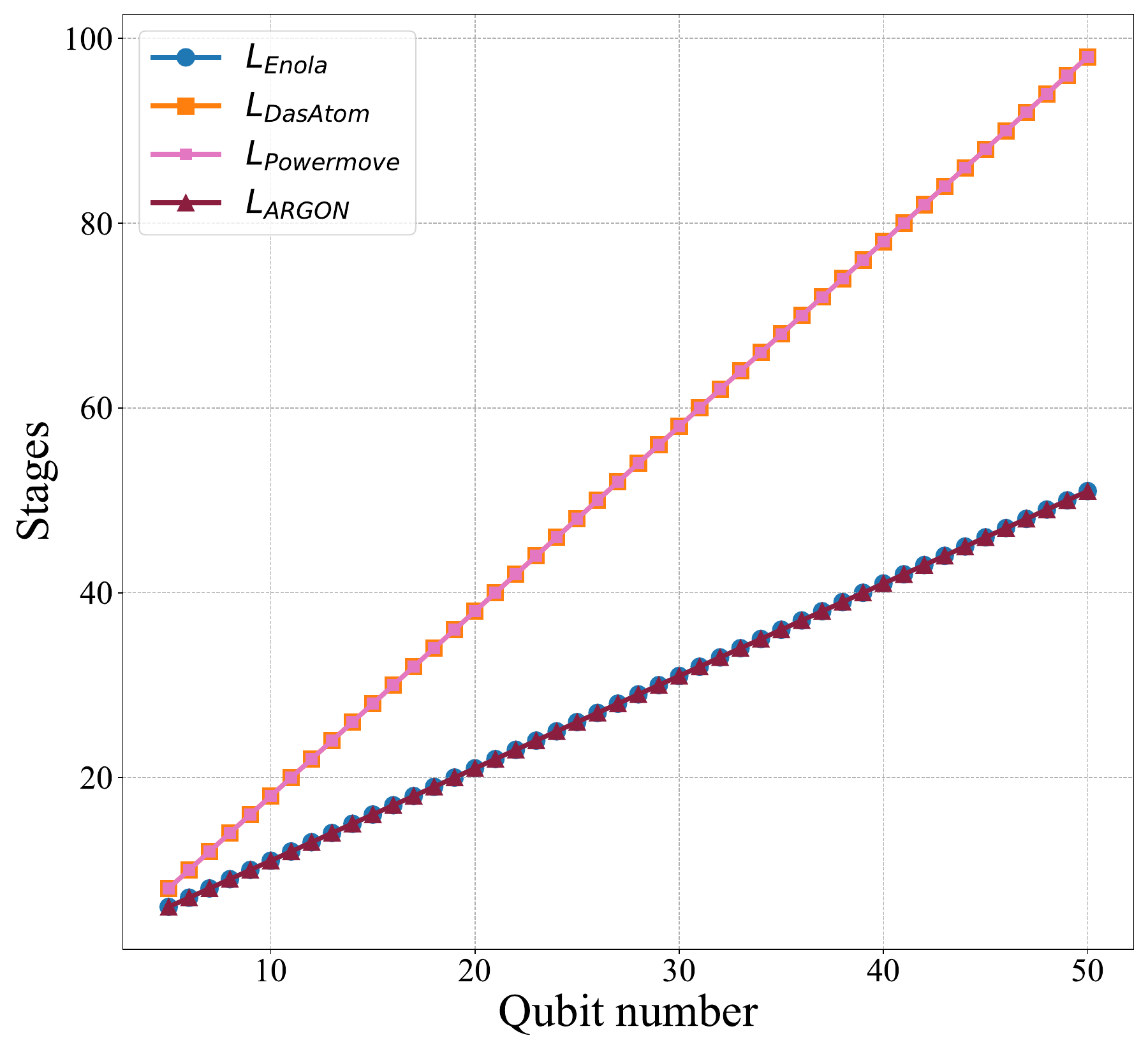}
		\caption{W-state: Rydberg stages}
	\end{subfigure}
	\hfill
	\begin{subfigure}{0.19\textwidth}
		\centering
		\includegraphics[width=\linewidth]{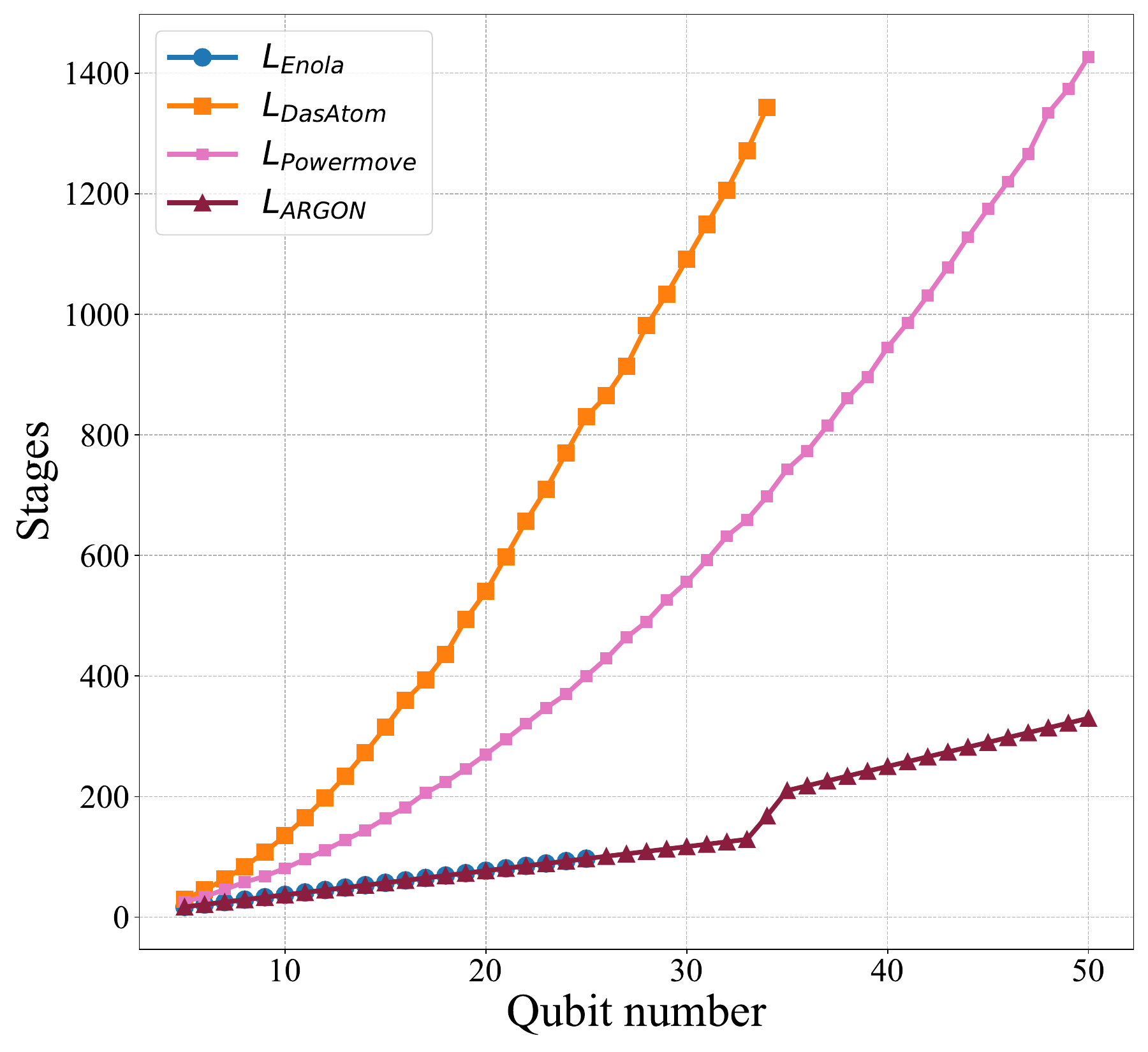}
		\caption{Random gates: Rydberg stages}
	\end{subfigure}
	\hfill
	\begin{subfigure}{0.19\textwidth}
		\centering
		\includegraphics[width=\linewidth]{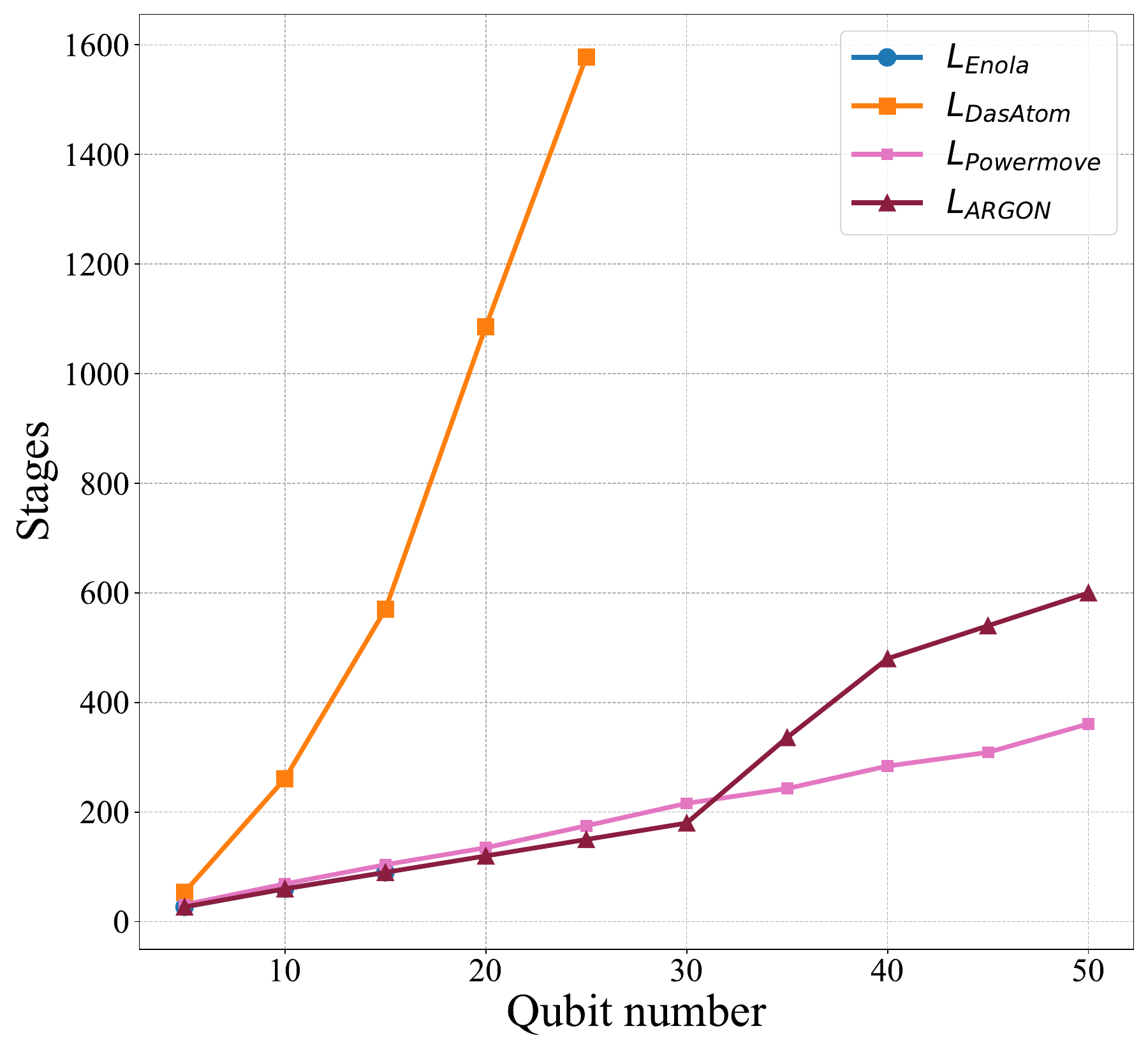}
		\caption{QV: Rydberg stages}
	\end{subfigure}
	\hfill
	\begin{subfigure}{0.19\textwidth}
		\centering
		\includegraphics[width=\linewidth]{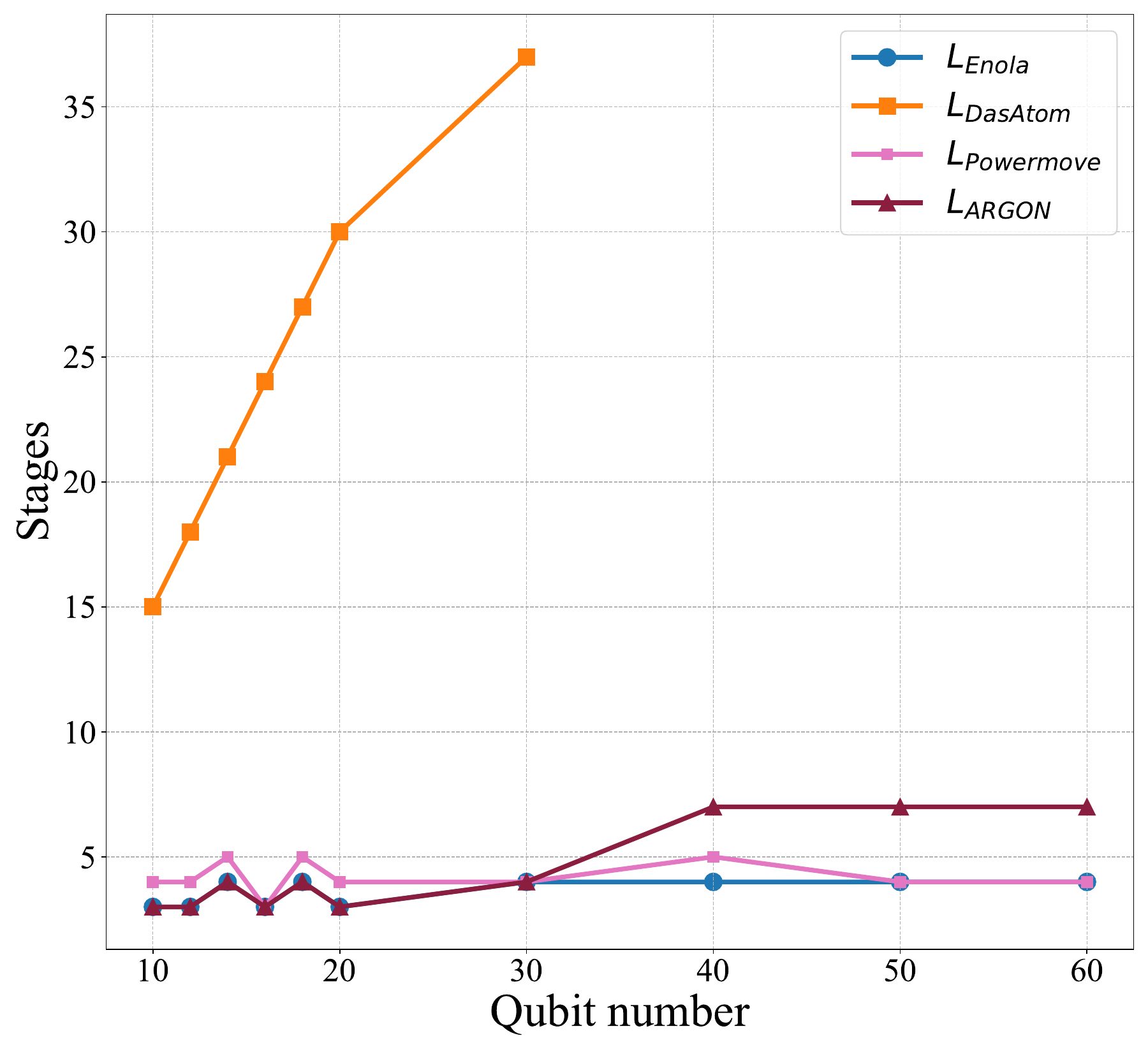}
		\caption{3-Regular: Rydberg stages}
	\end{subfigure}
	\hfill
	\begin{subfigure}{0.19\textwidth}
		\centering
		\includegraphics[width=\linewidth]{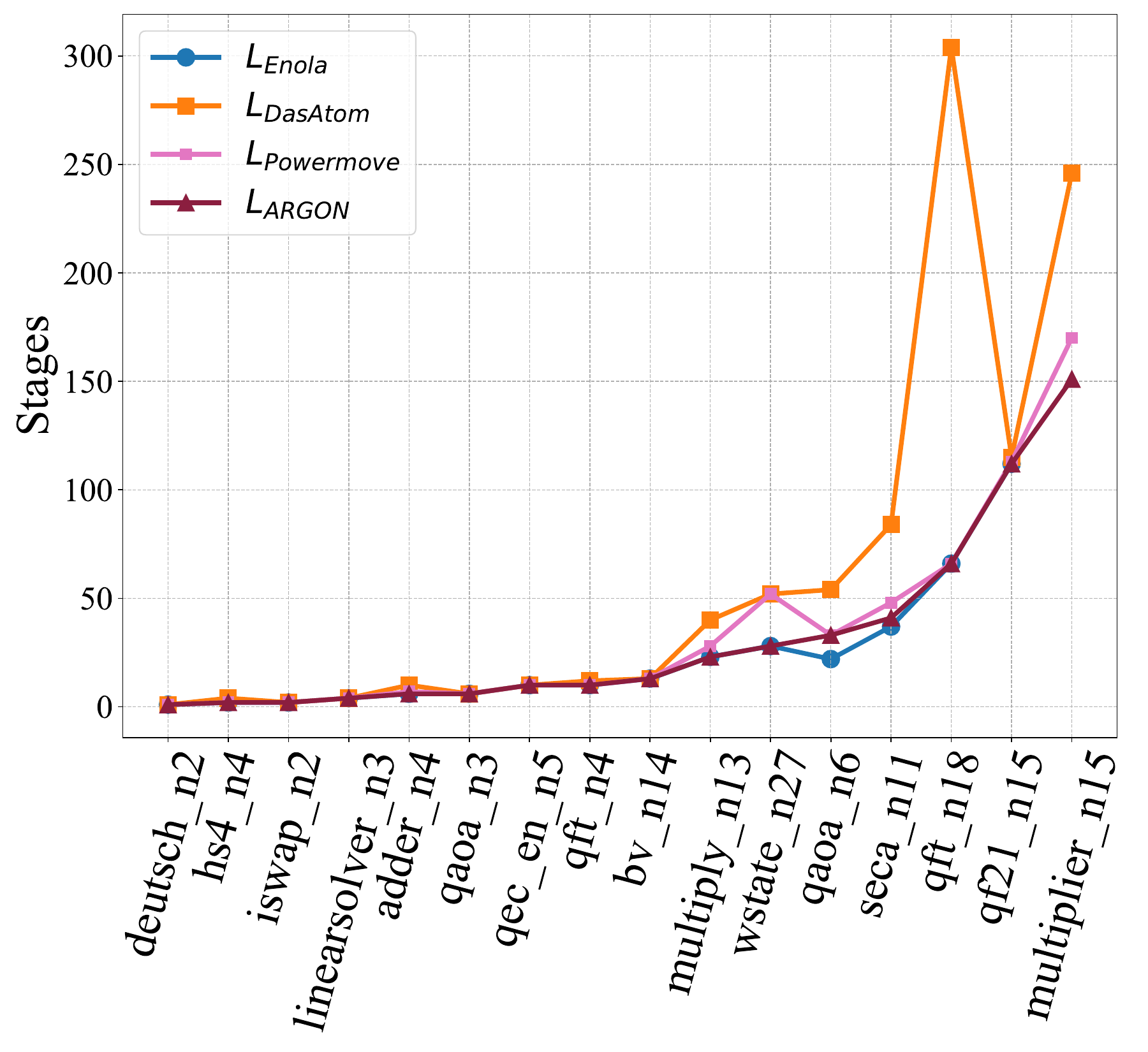}
		\caption{QASMbench: Rydberg stages}
	\end{subfigure}
	
	\caption{Experimental results on Enola, DasAtom, PowerMove, and ARGON. The first row $(a\sim d)$ compares compilation time. The second row $(e\sim h)$ illustrates circuit fidelity. The third row $(i\sim l)$ displays the number of Rydberg stages. Data points exceeding the time limit of 10{,}000\,s are omitted from the plots.}
	\label{fig:mqt_results_full}
\end{figure*}

\begin{table}[htbp]
	\centering
	\small
	\addtolength{\tabcolsep}{-2.5pt}
	\caption{Comparison of average compilation time (in seconds) across different benchmarks.}
	\label{table:time_avg}
	\begin{tabular*}{\linewidth}{l @{\extracolsep{\fill}} r r r r}
		\toprule
		\textbf{Benchmarks} & \textbf{ARGON} & \textbf{DasAtom~\cite{Huang2025}} & \textbf{PowerMove~\cite{Ruan2025}} & \textbf{Enola~\cite{Tan2025}}\\
		\midrule
		W-state        & \textbf{0.04}  & 36.9        & 58.4        & 295.4\\
		Random gates   & \textbf{1.1}   & $>3939.7$   & 989.9       & $>7156.4$\\
		QV             & \textbf{2.2}   & $>5397.1$   & 1338.3      & $>8342.5$\\
		3-Regular      & \textbf{0.02}  & $>3003.7$   & 32.7        & 43.4\\
		QASMBench      & \textbf{0.05}  & 0.05        & 96.3        & $>942.3$\\
		Total          & \textbf{0.68}  & $>2475.5$   & 503.1       & $>3356.0$\\
		\bottomrule
	\end{tabular*}
\end{table}

To empirically validate our spatiotemporal decoupling paradigm, this section evaluates the end-to-end system performance of ARGON against state-of-the-art baselines. We first analyze compilation efficiency to demonstrate the scalability of our framework, followed by an assessment of execution fidelity and Rydberg stage reduction to confirm its microarchitectural robustness. For the predictive policy, empirical profiling identified a lookahead window of $K=4$ as the optimal latency-accuracy trade-off.

\textbf{Compilation Efficiency and Scalability.} 
A primary objective of ARGON is to provide rapid compilation responsiveness for large-scale quantum processors. As summarized in Table~\ref{table:time_avg} and illustrated in Fig.~\ref{fig:mqt_results_full}, ARGON exhibits an average execution time of 0.68 seconds and completes all evaluated tasks within 10 seconds. Specifically, ARGON consistently achieves sub-second latency across the W-state, 3-regular, and QASMBench suites. For denser, high-load circuits like Random gates and Quantum Volume, ARGON maintains its efficiency, realizing an average speedup of over $600\times$ compared to the baselines. In contrast, DasAtom and Enola frequently trigger the $10^4$-second timeout threshold on these dense workloads. While PowerMove successfully avoids timeouts, its compilation latency on large-scale Random and QV circuits still approaches 2,000 seconds. This empirical bottleneck across existing coupled heuristics validates our core motivation: scalable compilation requires structurally decoupling static spatial parallelism from routing. Ultimately, the latency data substantiates that offloading spatial constraint resolution to an offline layout library effectively translates an intractable combinatorial search space into an efficient, manageable neural inference task, successfully breaking the scalability wall.

\textbf{Fidelity and Rydberg Stage Reduction.}

ARGON maintains strong execution fidelity as circuit scale increases. As illustrated in Fig.~\ref{fig:mqt_results_full}, ARGON achieves an approximate two-order-of-magnitude fidelity improvement over the baselines on the 50-qubit W-state circuit. For the 25-qubit Random gate benchmark, ARGON delivers fidelity gains of over 8 orders of magnitude compared to PowerMove, and over 10 orders of magnitude against Enola and DasAtom. On the practical QASMBench suite, ARGON attains an average physical fidelity of 0.575, outperforming PowerMove (0.559), DasAtom (0.562), and Enola (0.568). ARGON also demonstrates stable fidelity improvements across the 3-Regular suite. On the QV benchmarks, ARGON consistently outperforms DasAtom and Enola. For a small subset of larger QV circuits where PowerMove exhibits higher fidelity, the discrepancy stems from differing microarchitectural assumptions: PowerMove does not impose a strict upper bound on hardware spatial parallelism, effectively permitting unconstrained gate concurrency within a Rydberg stage. In contrast, ARGON enforces physical hardware constraints, yielding a more realistic compilation schedule.

These empirical results validate the architectural design. The offline layout library prevents the Rydberg stage inflation characteristic of localized heuristics, while the predictive placement policy provides the topological foresight needed to mitigate routing congestion and reduce idle decoherence.

Having established the system-level advantages of ARGON in both latency and fidelity, we next deconstruct these macro-level metrics.

\subsection{Compilation Time Breakdown}

\begin{figure}[htbp]
	\centering
	\begin{subfigure}{\linewidth} 
		\centering
		\includegraphics[width=0.95\linewidth]{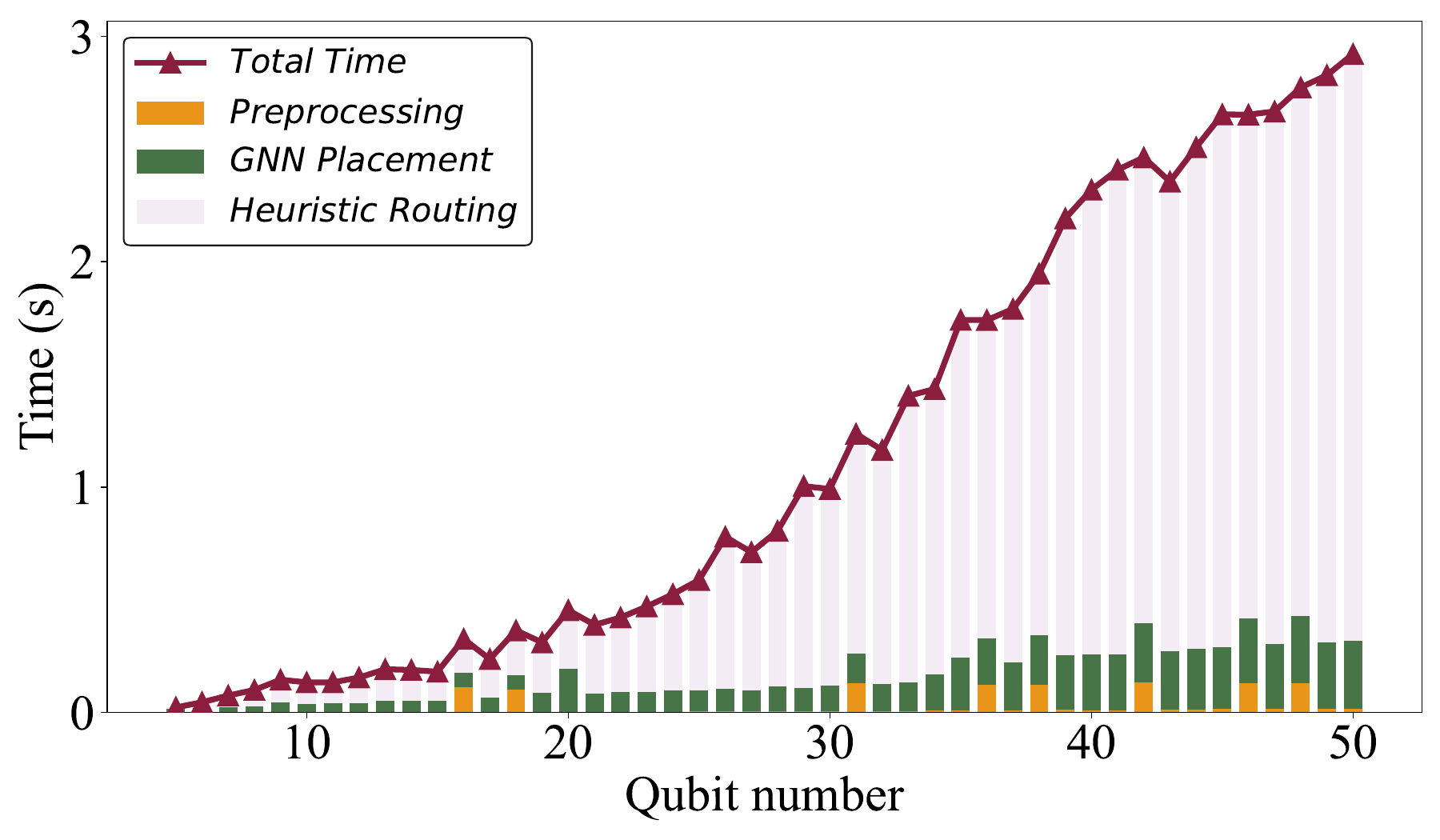} 
		\caption{Compilation time vs. number of qubits.}
		\label{fig:time_breakdown_q}
	\end{subfigure}
	\begin{subfigure}{\linewidth}
		\centering
		\includegraphics[width=0.95\linewidth]{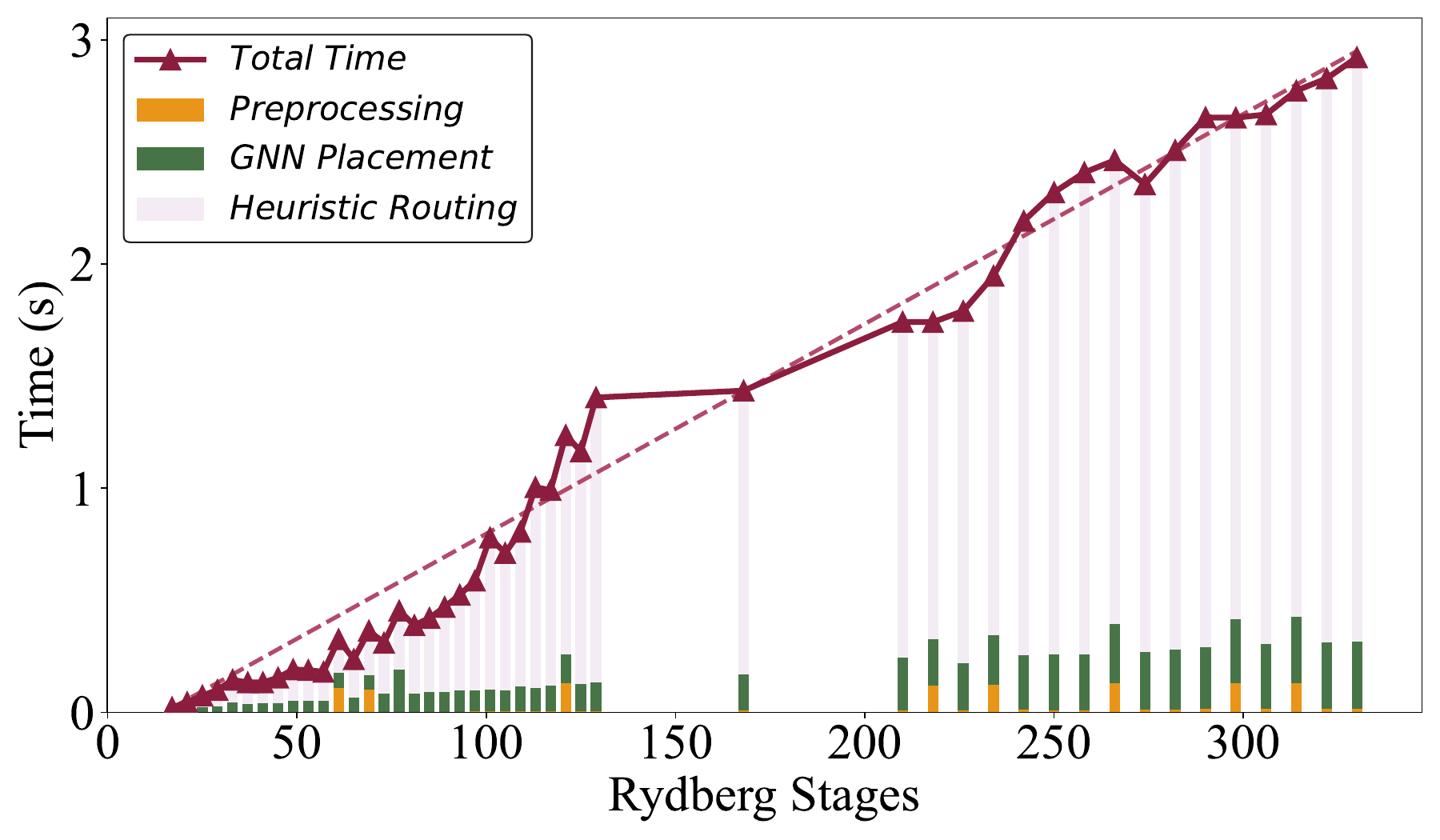}
		\caption{Compilation time vs. number of Rydberg stages.}
		\label{fig:time_breakdown_rstages}
	\end{subfigure}
	\caption{Execution time breakdown of ARGON evaluated on the Random gates benchmark. The total compilation time is decomposed into preprocessing, GNN placement, and heuristic routing.}
	\label{fig:time_breakdown}
\end{figure}

To analyze the compilation latency established in the previous section, we conduct a assessment of the execution time within the ARGON framework on the Random gates benchmark. As depicted in Fig.~\ref{fig:time_breakdown_q}, the lightweight heuristic routing process constitutes the dominant fraction of the compilation time. The combined latency of spatial library lookup, graph feature preprocessing, and GNN inference remains robustly bounded below 0.5 seconds even for dense 50-qubit workloads. Furthermore, Fig.~\ref{fig:time_breakdown_rstages} illustrates that the total compilation time scales linearly with the number of Rydberg stages. This linear complexity confirms the scalability of the backend routing phase, ensuring that ARGON maintains rapid execution (under 3 seconds) even for deep circuits requiring over 300 global stages.

These results validate ARGON's spatiotemporal decoupling paradigm. By shifting the computationally expensive spatial verification and layout generation to the offline phase, the compilation process is reduced to graph preprocessing, a neural forward pass, and lightweight routing heuristics, thereby avoiding the scheduling bottlenecks of conventional approaches.

\subsection{Compilation Fidelity Attribution}

\begin{figure}[htbp]
	\centering
	\begin{subfigure}{0.22\textwidth}
		\centering
		\includegraphics[width=\linewidth]{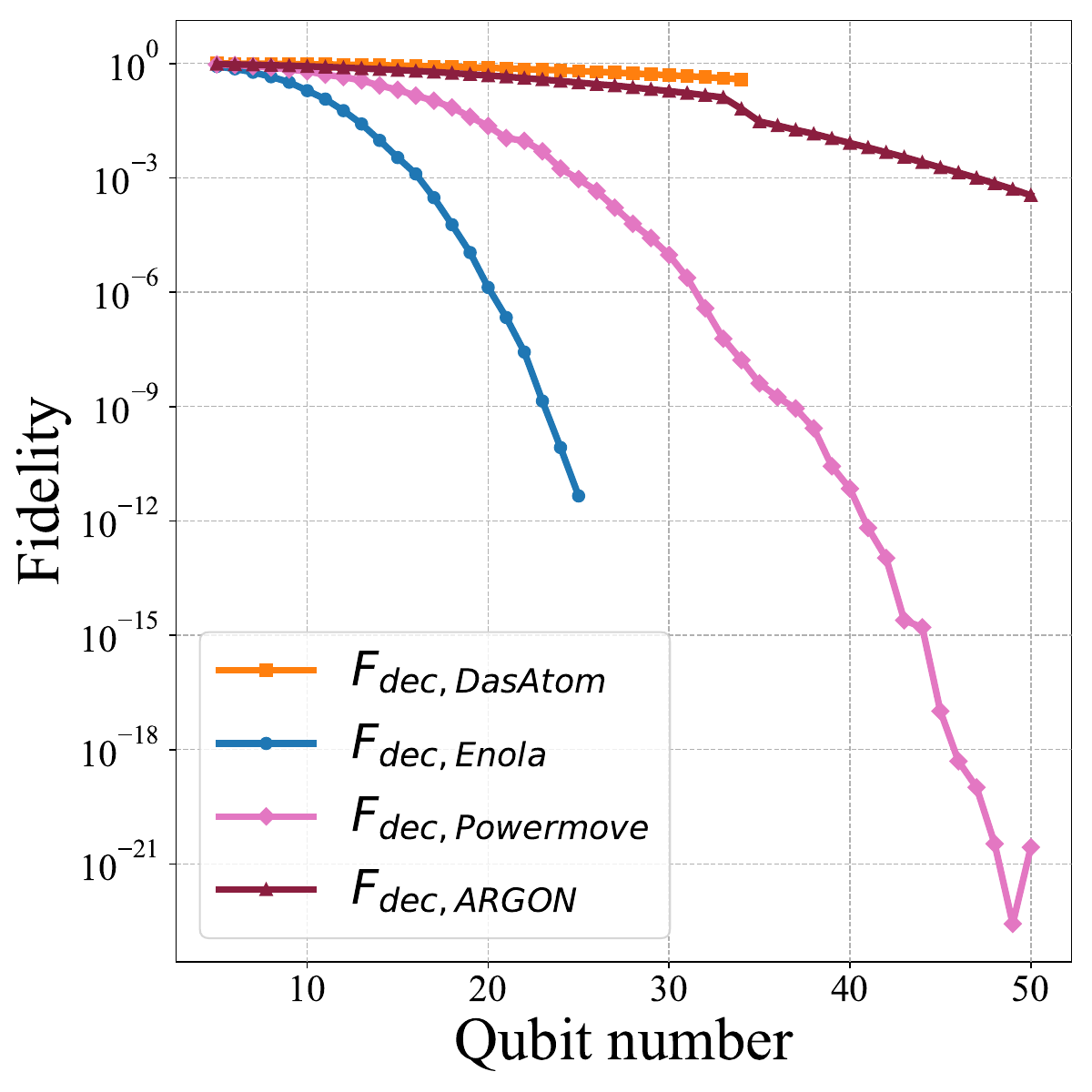}
		\caption{Decoherence fidelity ($F_{\mathrm{dec}}$)}
		\label{fig:fid_breakd_fdec}
	\end{subfigure}
	\hfill
	\begin{subfigure}{0.22\textwidth}
		\centering
		\includegraphics[width=\linewidth]{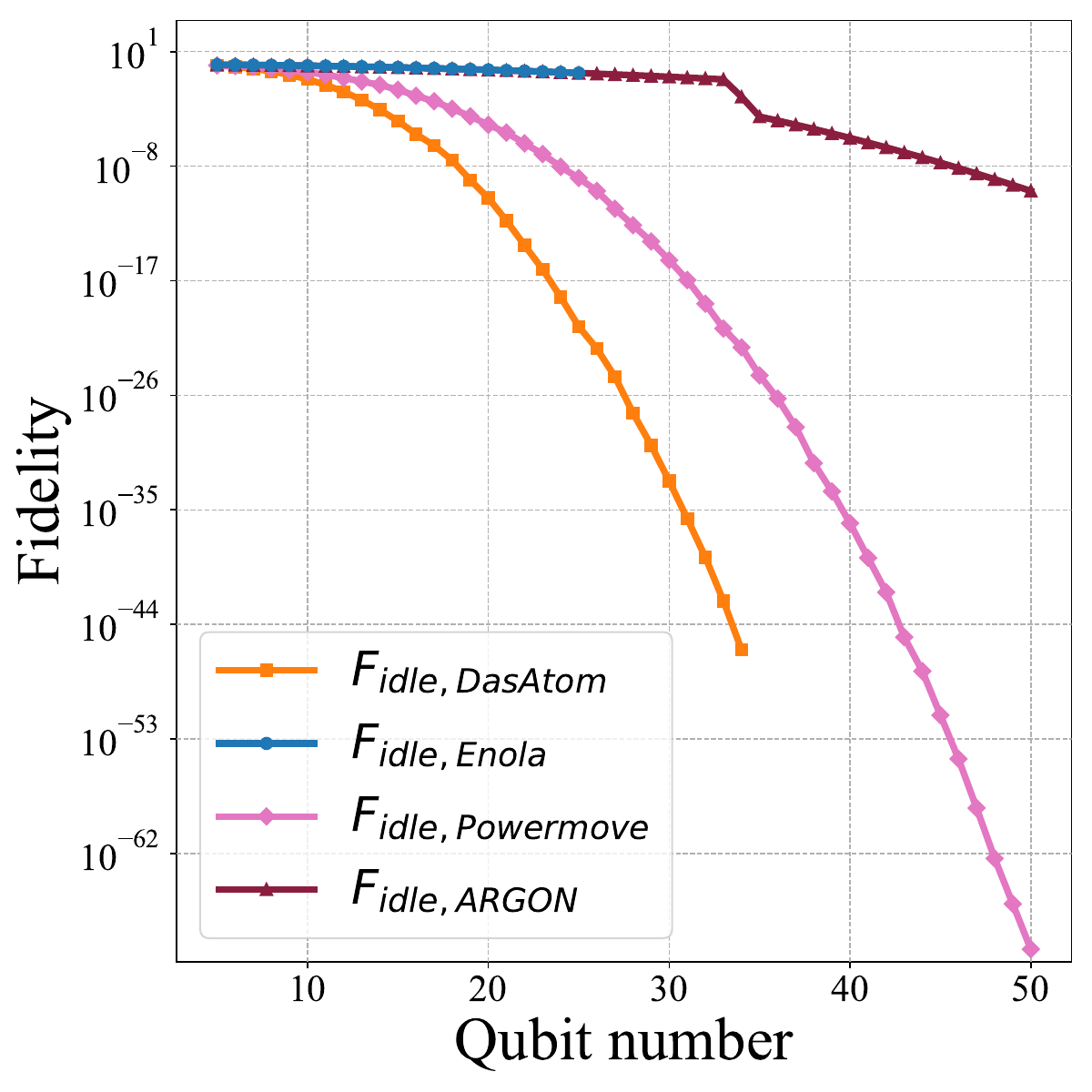}
		\caption{Idle qubit fidelity ($F_{\mathrm{idle}}$)}
		\label{fig:fid_breakd_fidle}
	\end{subfigure}
	
	\begin{subfigure}{0.22\textwidth}
		\centering
		\includegraphics[width=\linewidth]{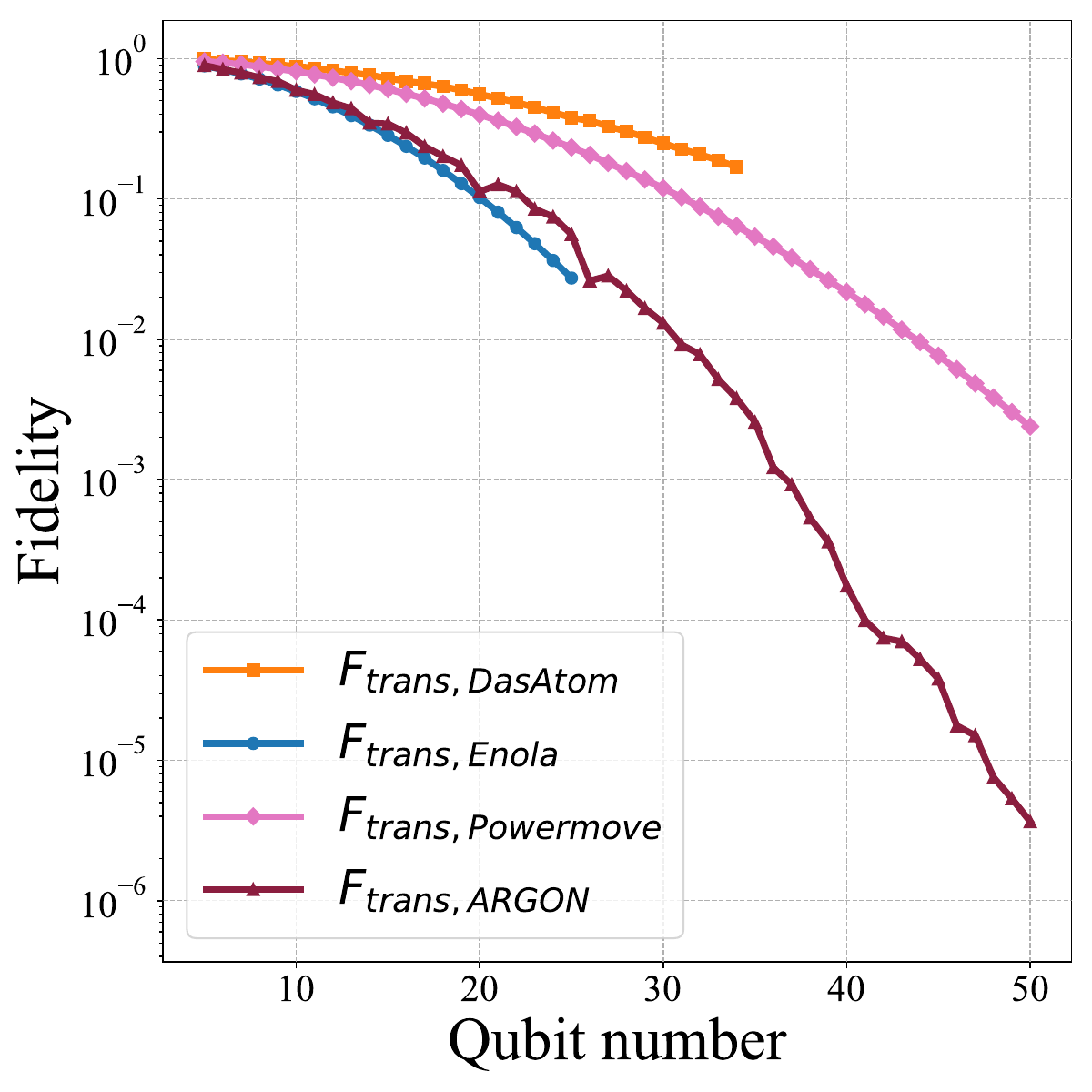}
		\caption{Atom transfer fidelity ($F_{\mathrm{trans}}$)}
		\label{fig:fid_breakd_ftrans}
	\end{subfigure}
	\hfill
	\begin{subfigure}{0.22\textwidth}
		\centering
		\includegraphics[width=\linewidth]{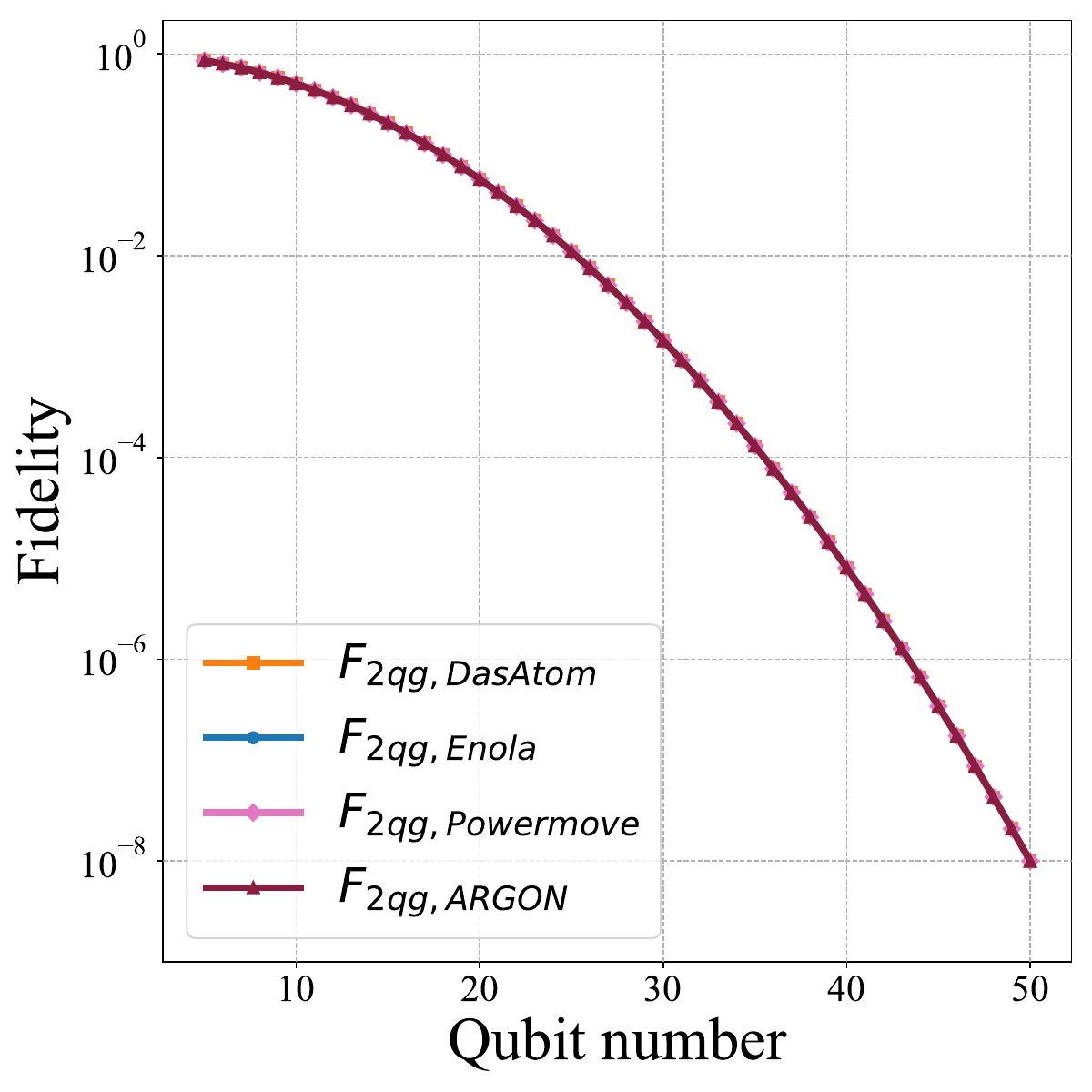}
		\caption{Two-qubit gate fidelity ($F_{\mathrm{2qg}}$)}
		\label{fig:fid_breakd_f2q}
	\end{subfigure}
	
	\caption{Detailed breakdown of fidelity components for different compilers as the number of qubits increases.}
	\label{fig:fidelity_breakdown}
\end{figure}

In this section, we analyze the physical execution costs to understand how different microarchitectural factors contribute to system-level fidelity. As illustrated in Fig.~\ref{fig:fidelity_breakdown}, the end-to-end fidelity is decomposed into four distinct components.

The results reveal different scaling characteristics across the evaluated frameworks. As shown in Fig.~\ref{fig:fid_breakd_fdec}, DasAtom achieves the highest movement-stage fidelity, followed closely by ARGON. DasAtom benefits from fewer movement stages enabled by subcircuit merging, whereas ARGON reduces inter-stage movement latency by selecting suitable layouts through GNN-based prediction. For idle-qubit fidelity (Fig.~\ref{fig:fid_breakd_fidle}), ARGON and Enola exhibit the best performance. In ARGON, this advantage stems from SMT-based layout generation, which maximizes gate-level parallelism, while Enola achieves a similar effect through its discretized site organization. Regarding atom-transfer fidelity (Fig.~\ref{fig:fid_breakd_ftrans}), DasAtom performs favorably due to requiring fewer atom transfer operations. Finally, since two-qubit gate fidelity depends solely on the number of two-qubit gates, all methods exhibit identical performance in this component, as shown in Fig.~\ref{fig:fid_breakd_f2q}.

Taken together, these results highlight the benefit of ARGON's spatiotemporal decoupling paradigm. By resolving static spatial constraints offline, ARGON maximizes gate parallelism and minimizes idle-excitation penalties, while the GNN-guided predictive placement reduces inter-stage movement overhead and the associated decoherence. These complementary optimizations across spatial and temporal dimensions translate into consistently higher end-to-end fidelity across a broad range of circuit scales. 

\subsection{Compilation Ablation and Generalization Study}

\begin{figure*}[t]
	\centering
	\begin{subfigure}{0.19\textwidth}
		\centering
		\includegraphics[width=\linewidth]{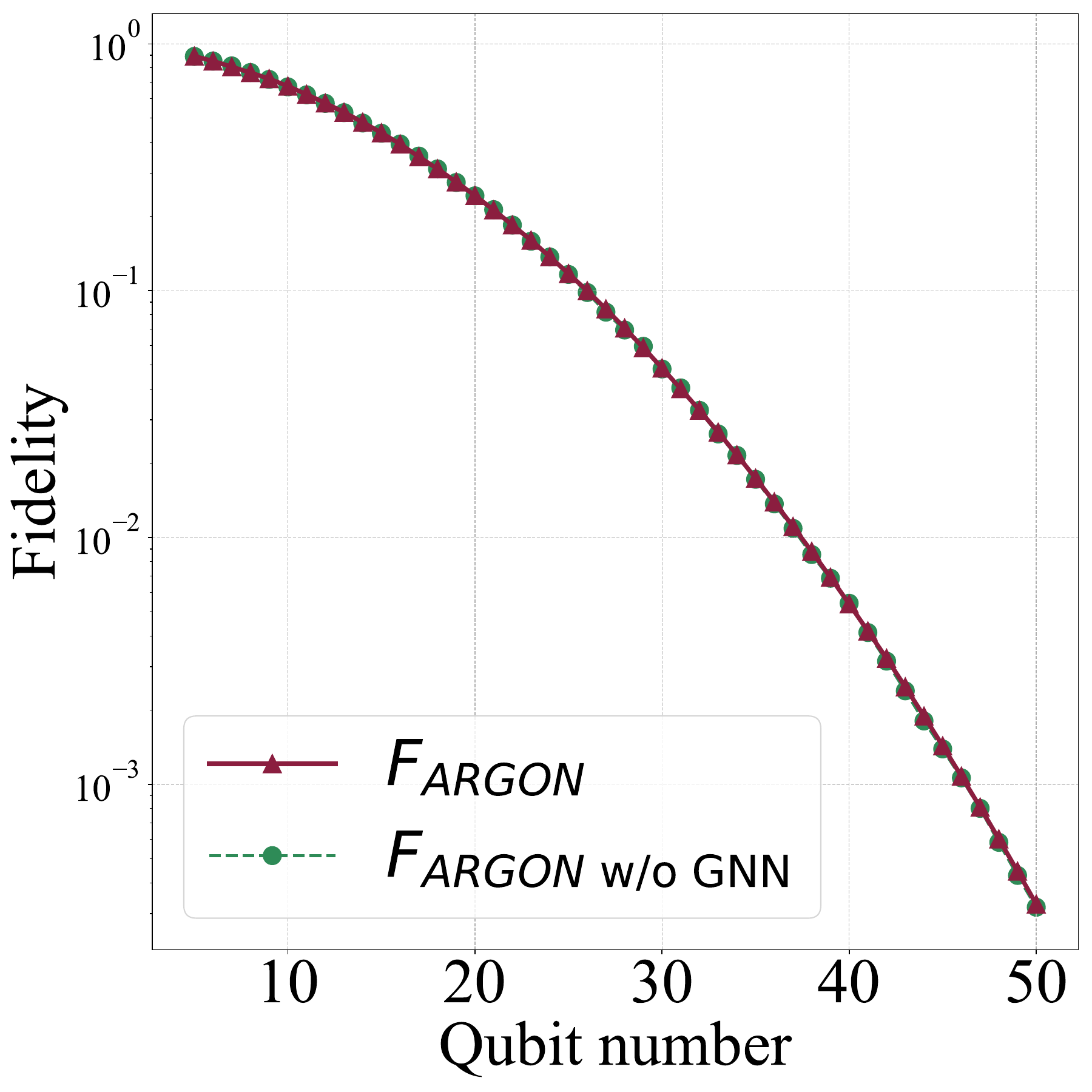}
		\caption{W-state}
		\label{fig:abl_wstate}
	\end{subfigure}
	\hfill
	\begin{subfigure}{0.19\textwidth}
		\centering
		\includegraphics[width=\linewidth]{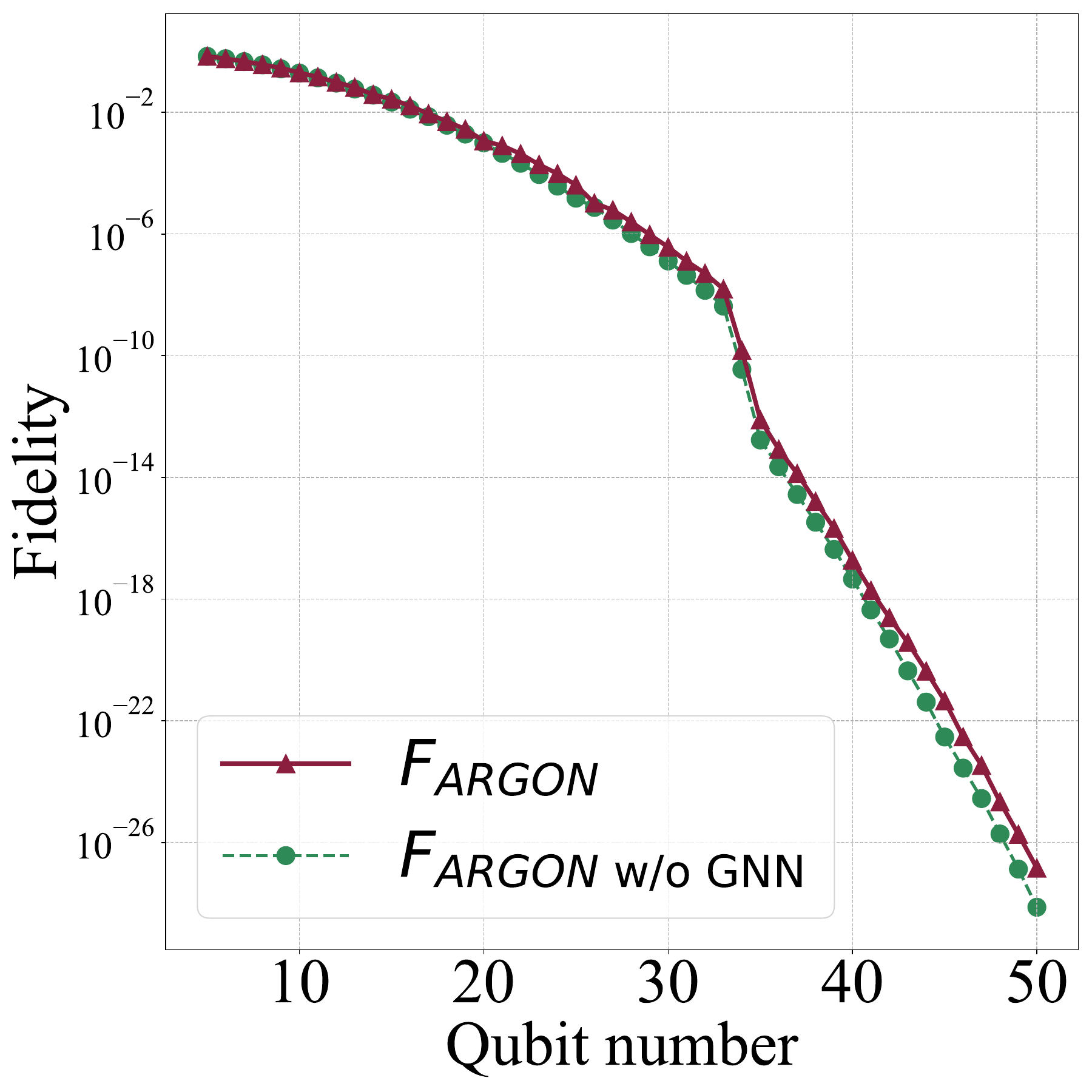}
		\caption{Random gates}
		\label{fig:abl_2qgate}
	\end{subfigure}
	\hfill
	\begin{subfigure}{0.19\textwidth}
		\centering
		\includegraphics[width=\linewidth]{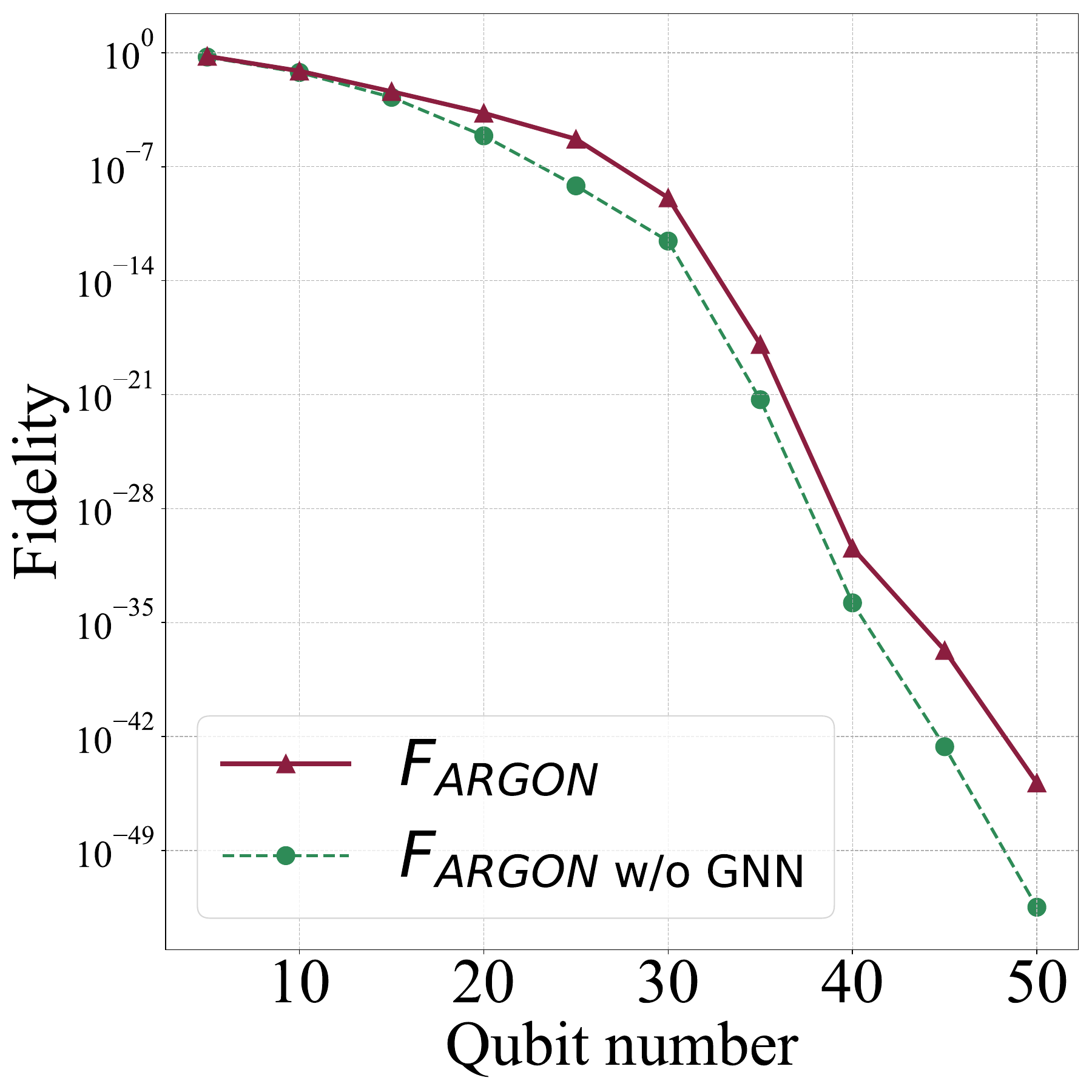}
		\caption{QV}
		\label{fig:abl_qv}
	\end{subfigure}
	\hfill
	\begin{subfigure}{0.19\textwidth}
		\centering
		\includegraphics[width=\linewidth]{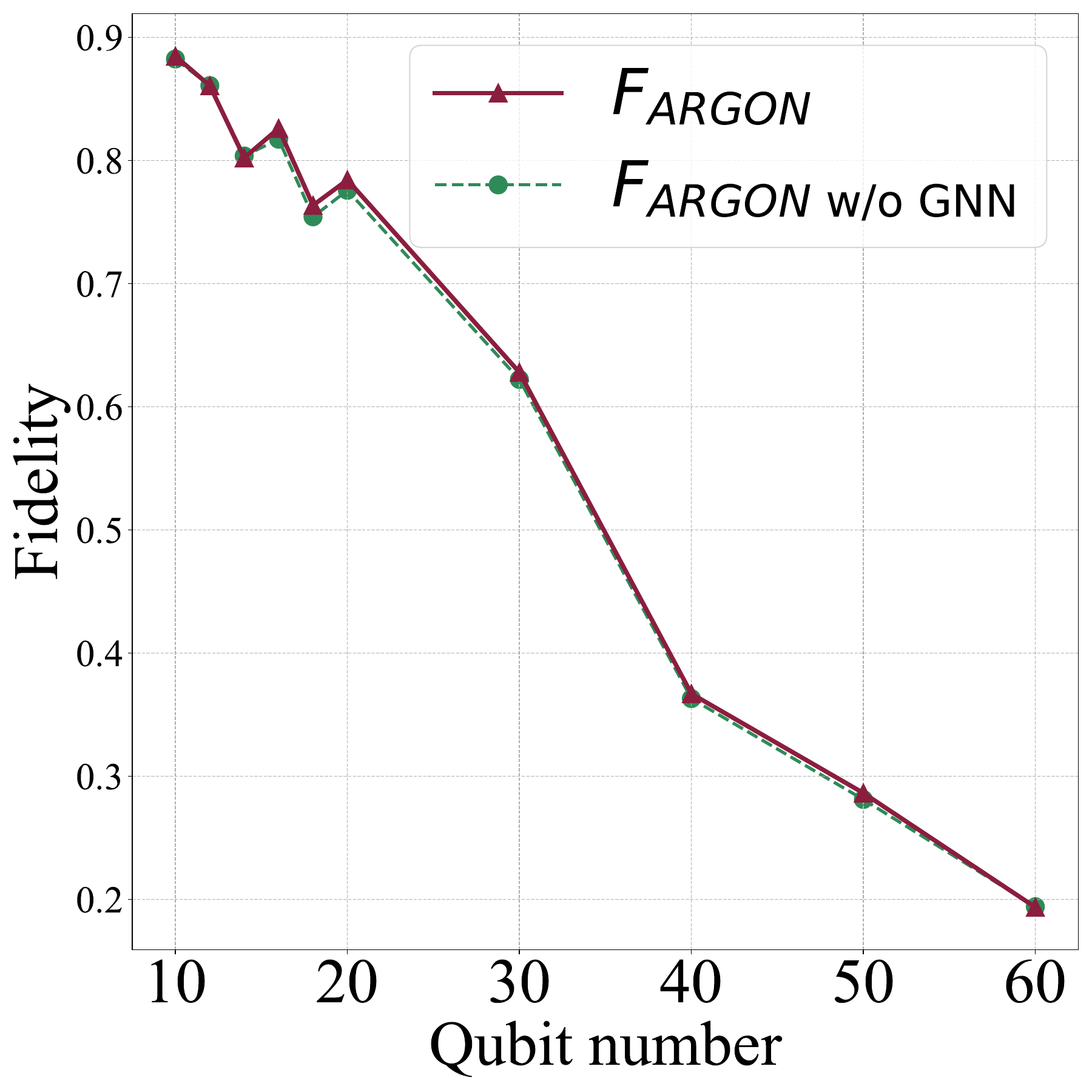}
		\caption{3-Regular}
		\label{fig:abl_3regular}
	\end{subfigure}
	\hfill
	\begin{subfigure}{0.19\textwidth}
		\centering
		\includegraphics[width=\linewidth]{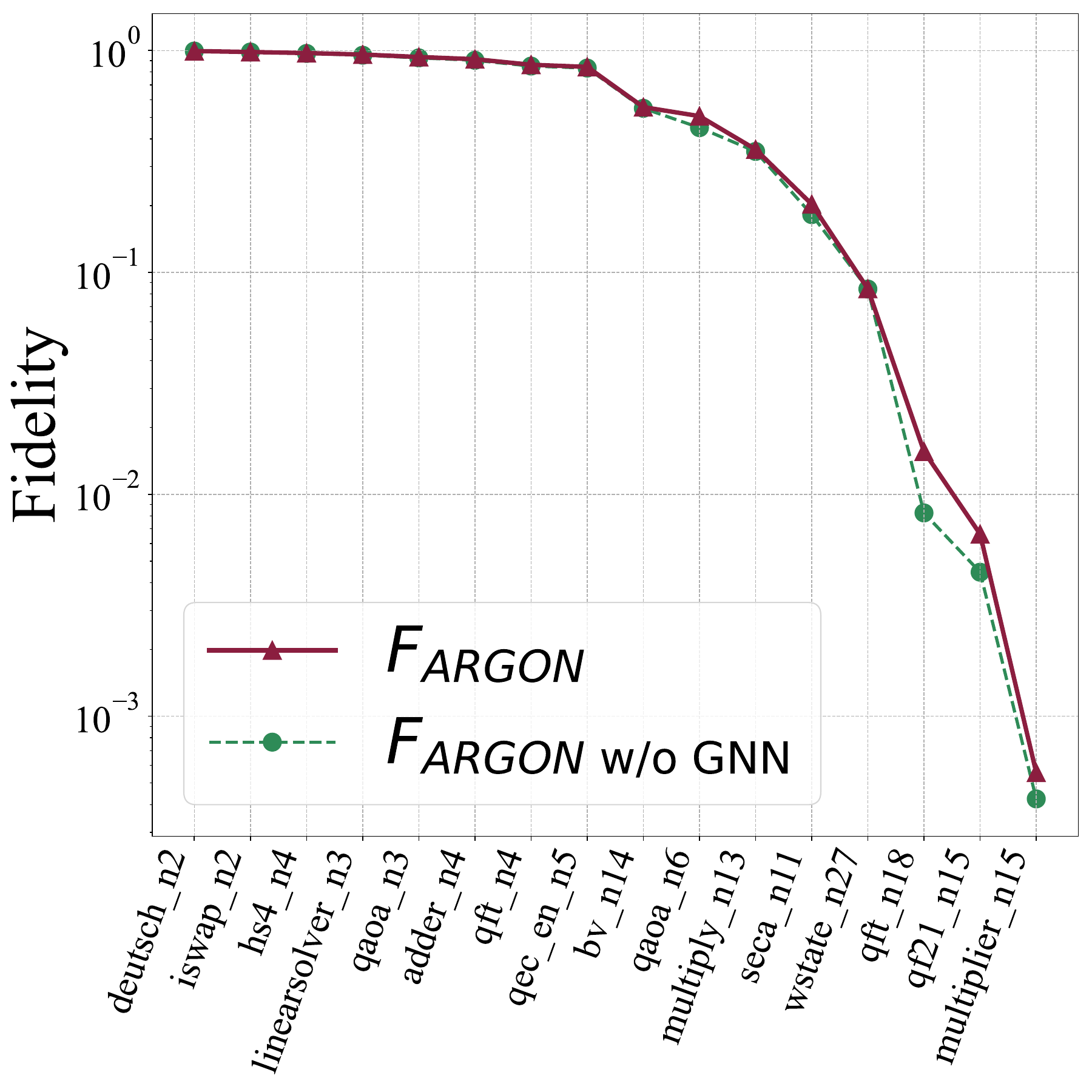}
		\caption{QASMbench}
		\label{fig:abl_qasmbench}
	\end{subfigure}
	
	\caption{Ablation study evaluating the efficacy of the GNN-guided predictive policy. $F_{\mathrm{ARGON}}$ denotes ARGON with the full model and $F_{\mathrm{ARGON\ w/o\ GNN}}$ denotes the baseline without the GNN module.}
	\label{fig:ablation_gnn}
\end{figure*}

\begin{figure}[t]
	\centering
	\begin{subfigure}{0.23\textwidth}
		\centering
		\includegraphics[width=\linewidth]{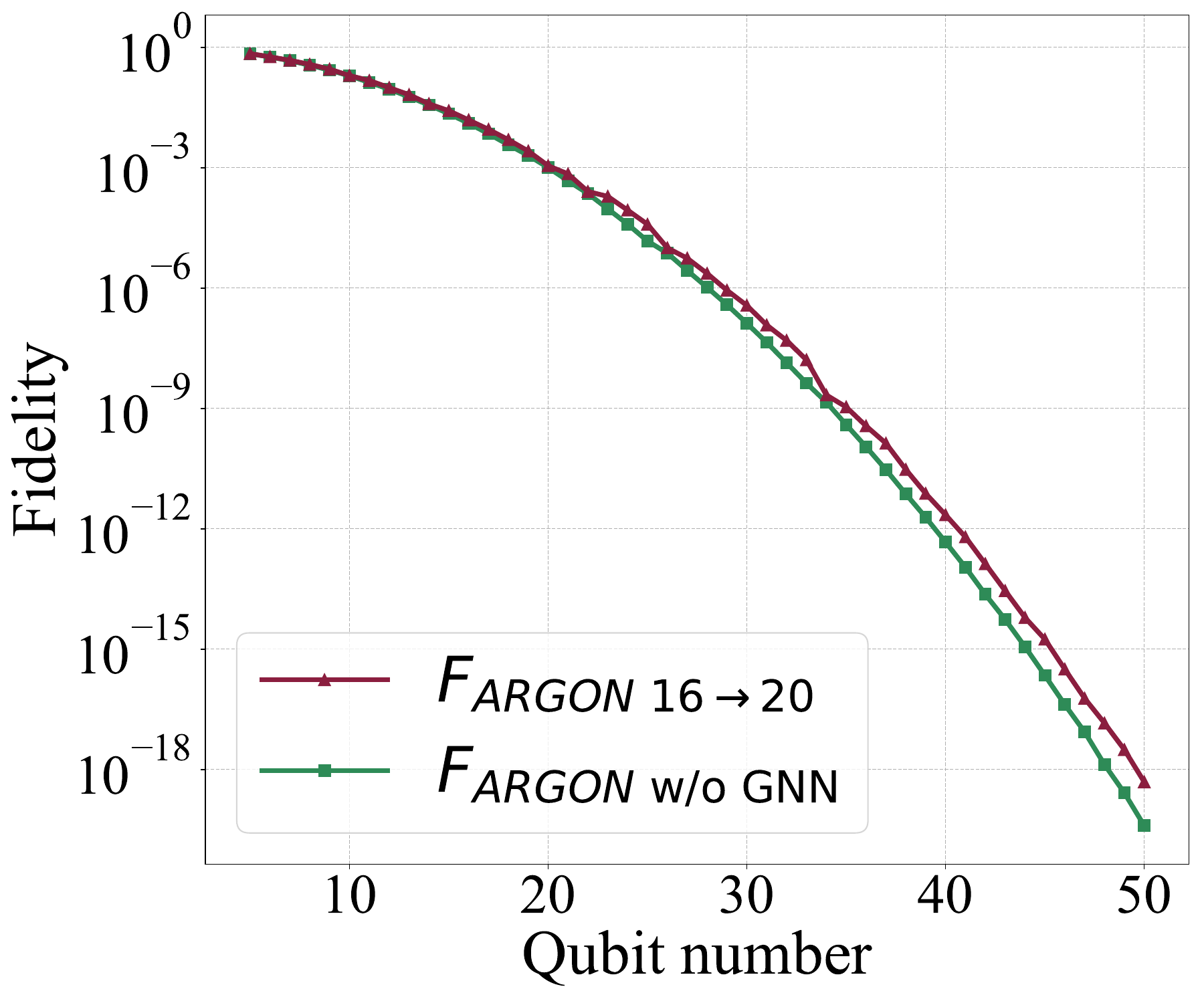}
		\caption{$20 \times 20$ array}
		\label{fig:zeroshot_20}
	\end{subfigure}
    \hfill
	\begin{subfigure}{0.23\textwidth}
		\centering
		\includegraphics[width=\linewidth]{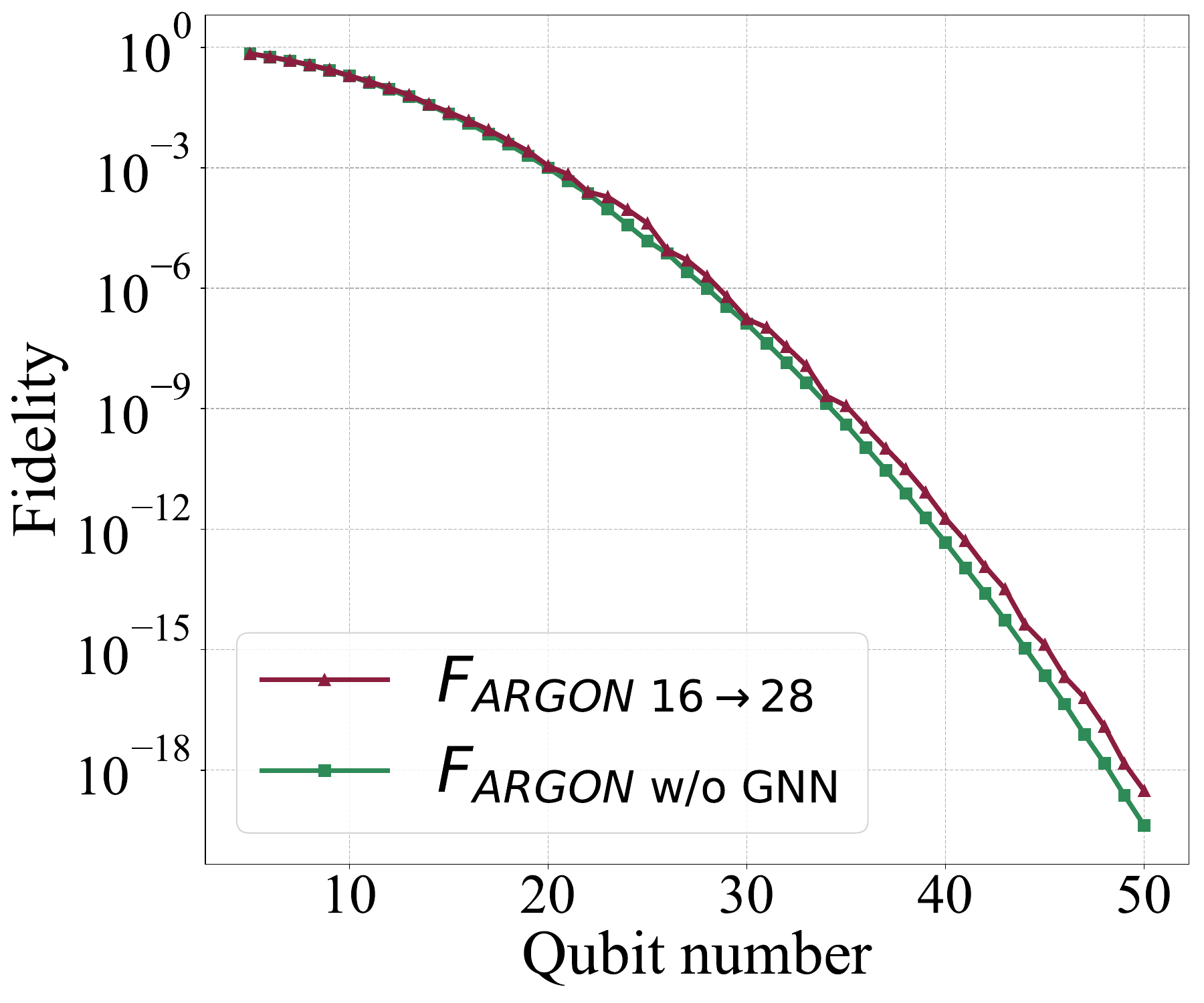}
		\caption{$28 \times 28$ array}
		\label{fig:zeroshot_28}
	\end{subfigure}
	\caption{Zero-shot generalization evaluation of the GNN-guided policy on the Random gates benchmark. The GNN module, trained exclusively on a $16 \times 16$ array, is directly applied to larger hardware topologies: (a) $20 \times 20$ and (b) $28 \times 28$ arrays.}
	\label{fig:zeroshot_generalization}
\end{figure}

To validate the driving mechanism behind the spatiotemporal decoupling paradigm, we evaluate the microarchitectural contribution and structural generalizability of the GNN. We first compare the full ARGON framework against a baseline that employs identical offline layout libraries but relies on randomized layout selection rather than neural prediction. As shown in Fig.~\ref{fig:ablation_gnn}, empirical evaluations on the Random gates benchmark reveal that substituting the neural policy with random selection results in a fidelity drop of up to an order of magnitude. This performance gap indicates that the offline layout library requires an intelligent selection mechanism to unlock its architectural potential.

To evaluate structural robustness across varying hardware dimensions, we deploy the predictive model trained exclusively on a $16 \times 16$ array directly onto larger $20 \times 20$ and $28 \times 28$ topologies. As illustrated in Fig.~\ref{fig:zeroshot_generalization}, the pre-trained model consistently outperforms the random selection baseline across all expanded hardware scales. The fidelity gap visibly widens as circuit size increases, confirming the sustained effectiveness of the predictive routing policy under expanded topological constraints.

These combined ablation and zero-shot evaluations empirically establish the physical scaling mechanism of the framework. The offline-generated layouts function as universal microarchitectural primitives. Because physical exclusion limits dictate local geometric neighborhoods rather than global grid boundaries, the learned composition logic inherently possesses layout scale-invariance. By learning to compose these discrete spatial templates, the neural policy ensures robust layout matching prior to execution and systematically tessellates local structural patterns across scaled architectures without requiring any offline re-computation or library expansion.

\subsection{Scalability of the Offline Library and Overhead Analysis}
To assess the practical overhead of the spatiotemporal decoupling paradigm, we analyze the scalability of the offline layout library, its memory footprint, and its impact on the compile-time GNN evaluation.

\begin{figure}[t]
	\centering
	\begin{subfigure}{0.23\textwidth}
		\centering
		\includegraphics[width=\linewidth]{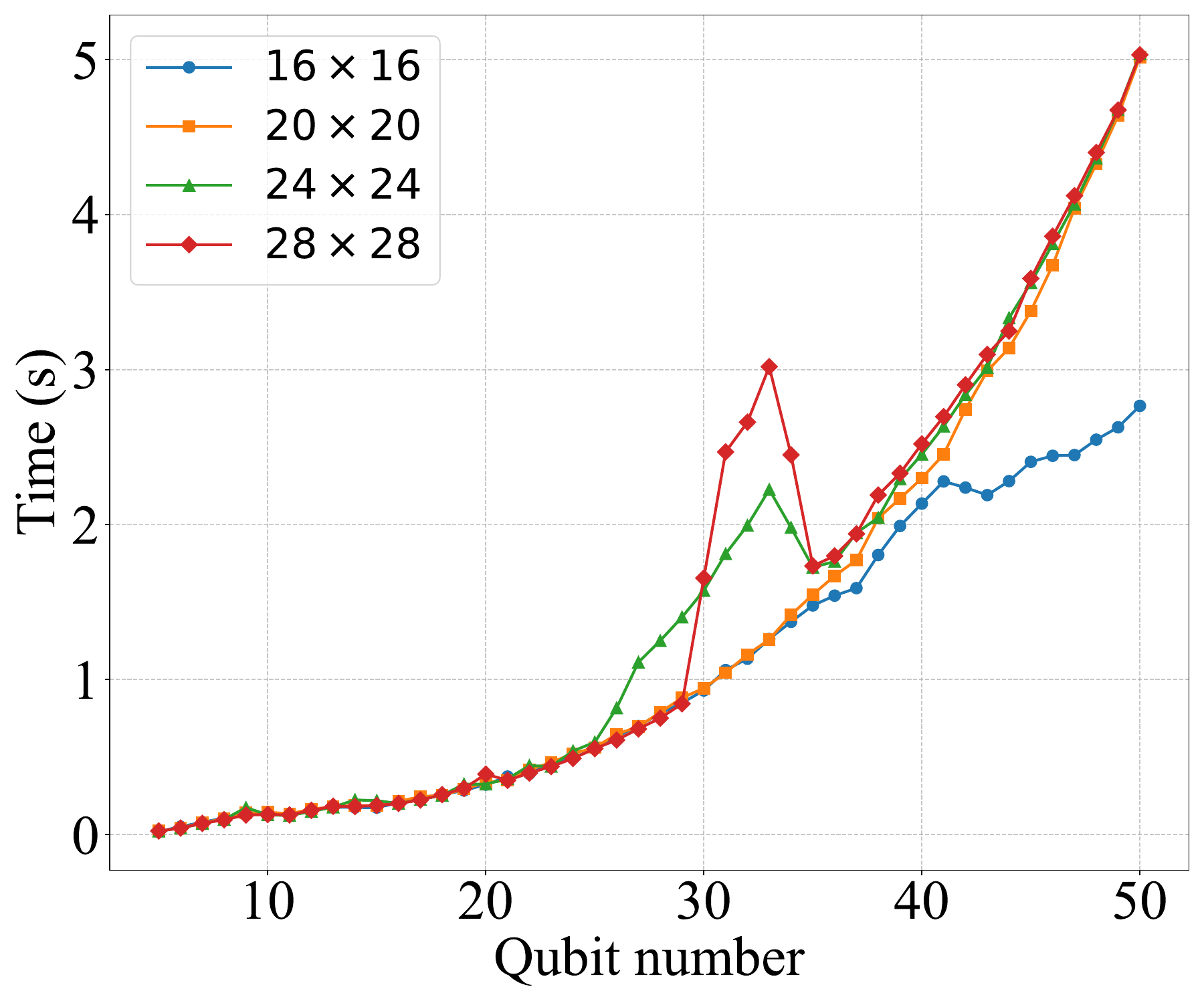}
		\caption{Hardware dimension scaling}
		\label{fig:scale_grid}
	\end{subfigure}
	\hfill
	\begin{subfigure}{0.23\textwidth}
		\centering
		\includegraphics[width=\linewidth]{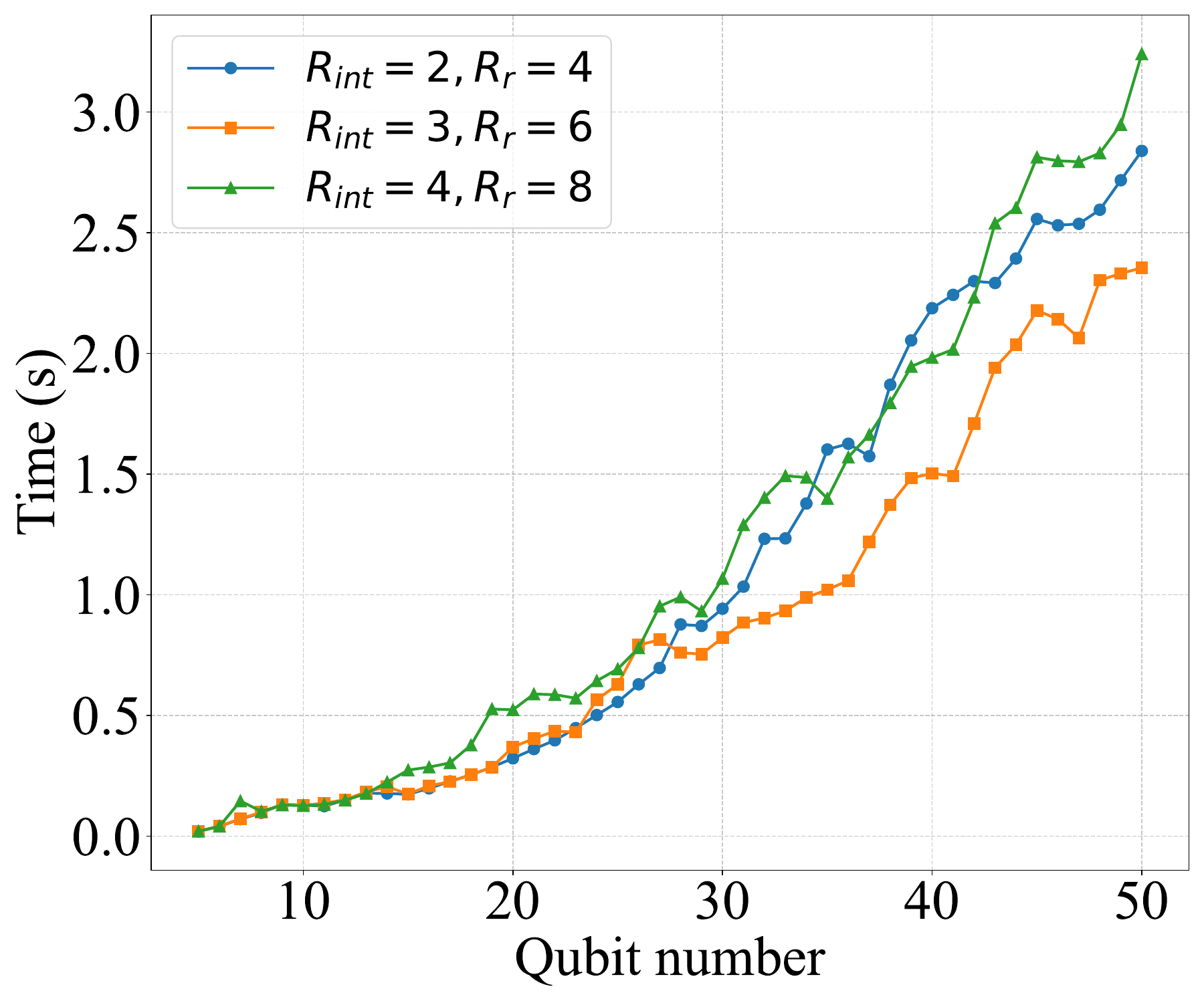}
		\caption{Interaction radius scaling}
		\label{fig:scale_radius}
	\end{subfigure}
	\caption{Evaluation of compile-time overhead. (a) Compilation time scaling across varying physical hardware grid dimensions (from $16 \times 16$ to $28 \times 28$). (b) Compilation time under expanding interaction ($R_{int}$) and restriction ($R_r$) radii constraints.}
	\label{fig:overhead_analysis}
\end{figure}

\textbf{Non-Recurring Engineering Overhead and Footprint.}
Because the offline SMT solver targets localized geometric constraints rather than global algorithmic logic, the number of unique layouts scales moderately with hardware dimensions. The template generation itself remains lightweight, requiring 3.5 s on a $10 \times 10$ grid, 12.1 s on a $16 \times 16$ grid, and 21.7 s on a $20 \times 20$ grid. The offline phase generates 104 layouts for a $16 \times 16$ grid, 128 layouts for a $20 \times 20$ grid, and 152 layouts for a $24 \times 24$ grid. The memory footprint for storing and evaluating these candidates remains modest: across all benchmarks, ARGON consumes at most $1.5$~GB of system RAM and under $0.5$~GB of GPU VRAM, enabling deployment on standard workstations. The one-time NRE cost for dataset generation and GNN training is approximately 3 hours and 30 minutes, respectively. Since this process depends only on hardware topology, it is a one-time amortized cost that does not impact daily compilation latency.

\textbf{Impact on Compilation Time.}
We further evaluated the impact of varying hardware parameters on compile-time overhead using the dense Random gates benchmark. Fig.~\ref{fig:scale_grid} demonstrates the compilation time as the physical hardware grid scales from $16 \times 16$ to $28 \times 28$. Although larger grids expand the discrete candidate pool and kinematic routing space, the compilation time scales efficiently, remaining below 5 seconds for Random gate circuits up to 50 qubits. Similarly, Fig.~\ref{fig:scale_radius} illustrates the evaluation overhead when expanding the interaction and restriction radii ($R_{int}$ and $R_r$). The compilation time exhibits robust scaling, staying under 3.5 seconds across all tested radii configurations for the 50-qubit Random circuits. These empirical results confirm that the decoupled offline library provides sufficient spatial diversity without introducing an exponential evaluation bottleneck.

\subsection{Comparison with Zoned-Architecture Heuristics}

\begin{figure}[t]
	\centering
		\begin{subfigure}{0.49\linewidth}
		\centering
		\includegraphics[width=\linewidth]{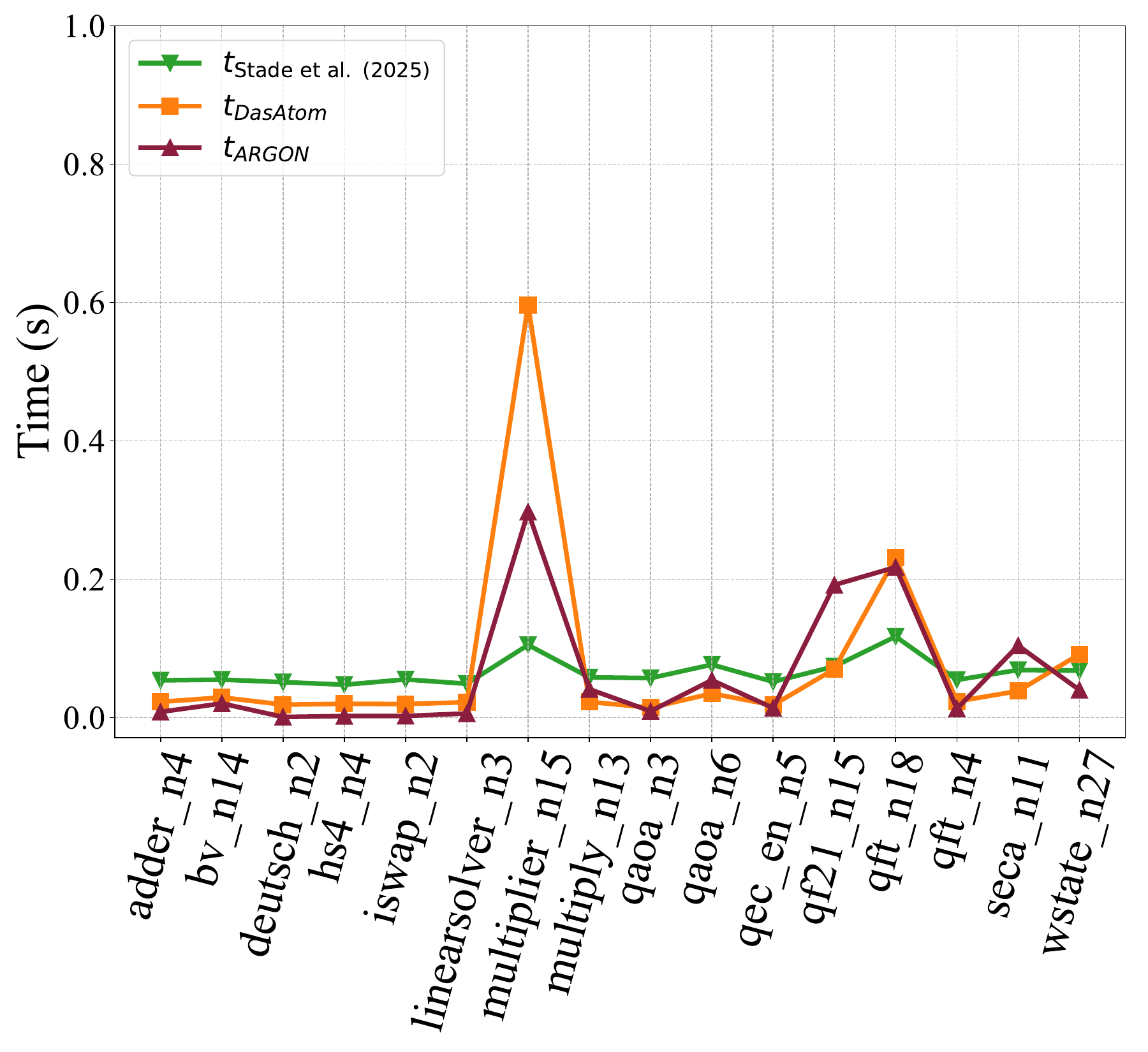}
		\caption{Compilation time ($t$)}
		\label{fig:zoned_time}
	\end{subfigure}
	\hfill
	\begin{subfigure}{0.49\linewidth}
		\centering
		\includegraphics[width=\linewidth]{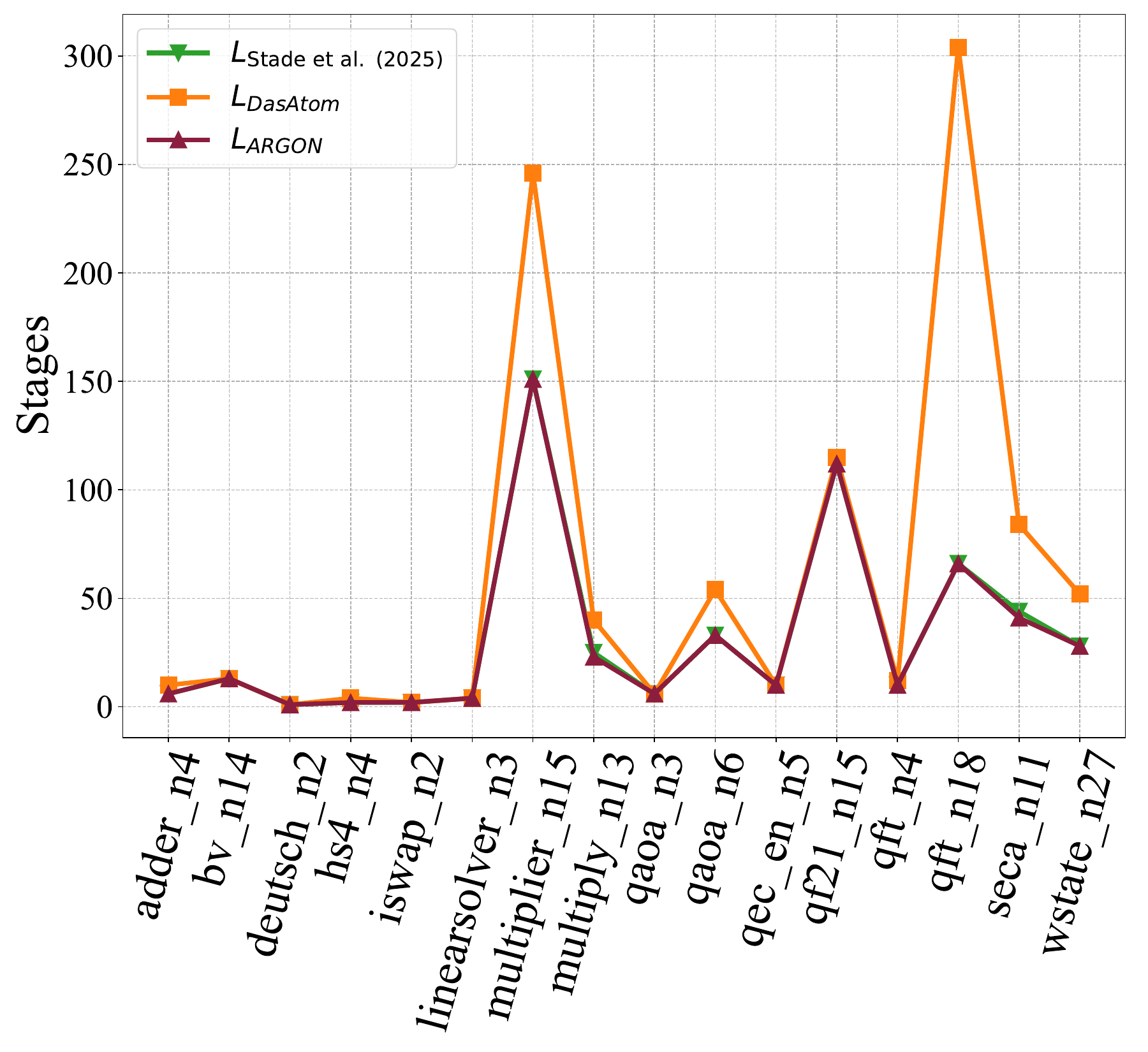}
		\caption{Rydberg stages ($L$)}
		\label{fig:zoned_stages}
	\end{subfigure}

	\caption{Performance comparison between ARGON, DasAtom, and the zoned-architecture heuristic by Stade et al.~\cite{Stade2025} on QASMBench.}
	\label{fig:zoned_cmp}
\end{figure}

To contextualize ARGON within the emerging trend of zoned architectures, we compare it against a recent A*-based placement heuristic by Stade et al.~\cite{Stade2025}. As shown in Fig.~\ref{fig:zoned_cmp}, all evaluated frameworks achieve rapid, sub-second compilation on the QASMBench suite. ARGON and Stade et al.~\cite{Stade2025} average 0.063s and 0.065s, respectively, both outperforming DasAtom (0.079s). While the heuristic by Stade et al. demonstrates impressive agility by efficiently optimizing immediate-layer routing, this localized focus can occasionally induce downstream congestion. Conversely, ARGON's GNN-based multi-layer foresight proactively mitigates these long-term bottlenecks, yielding fewer global Rydberg stages on specific circuits (e.g., \texttt{seca\_n11} and \texttt{multiply\_n13}).

We omit a direct comparison of physical execution fidelity for two practical reasons: first, Stade et al.~\cite{Stade2025} does not provide explicit hardware parameters or fidelity formulations; second, because their algorithm is structurally hardcoded to zoned architectures, applying our monolithic penalty model would introduce microarchitectural inconsistencies. Ultimately, this evaluation highlights a balanced algorithmic trade-off: localized heuristics excel in agile immediate-layer placement, whereas predictive data-driven models like ARGON provide robust long-term routing efficiency.

\section{Discussion and Conclusion} 

The architectural design of ARGON provides practical insights for the evolution of neutral atom systems. A key implication of the spatiotemporal decoupling paradigm is the scale invariance of offline templates. Because physical exclusion constraints define local geometric neighborhoods rather than global array dimensions, precomputed layouts act as universal microarchitectural primitives independent of system scale. This enables zero-shot generalization to larger hardware topologies and allows the template library and predictive routing engine to naturally serve as an efficient scheduler for dense entangling regions in future zoned architectures. Moreover, since these templates are governed by invariant physical rules rather than algorithmic logic, the resulting layout library constitutes a valuable microarchitectural dataset for broader data-driven explorations. While ARGON targets dense monolithic architectures, full zoned systems and FTQC stacks additionally require storage management and inter-zone traffic orchestration, where ARGON naturally serves as an entangling-region scheduling primitive within these broader system stacks.

In conclusion, compile-time spatiotemporal coupling creates a major scalability bottleneck for reconfigurable quantum processors. ARGON addresses this by shifting spatial constraint verification to an offline library. This explicit decoupling is a necessary architectural choice, as a joint spatiotemporal neural model would substantially enlarge the action space and struggle to jointly represent discrete geometric constraints and continuous routing trajectories. By ensuring spatial validity offline, the compile-time phase employs a GNN to proactively mitigate downstream routing conflicts. Compared with sequence models and standard MLPs, GNNs provide a suitable inductive bias for capturing both the invariant hardware interaction graph and the temporal circuit dependency graph. Experimental results confirm that ARGON transforms combinatorial search into an efficient inference problem, achieving rapid compilation and improved execution fidelity. By bridging physical kinematic constraints and global temporal scheduling, ARGON establishes a robust and scalable compilation foundation.

%\begin{acks}
%	This document is derived from previous conferences, in particular MICRO 2013, ASPLOS 2015, MICRO 2015-2025, ISCA 2025, as well as SIGARCH/TCCA's Recommended Best Practices for the Conference Reviewing Process. 
%\end{acks}

\clearpage

\bibliographystyle{ACM-Reference-Format}
\bibliography{ref}

\end{document}